\documentclass[acmsmall]{acmart}

\AtBeginDocument{%
  }

\setcopyright{acmlicensed}
\acmJournal{PACMHCI}
\acmYear{2024} \acmVolume{8} \acmNumber{CSCW2} \acmArticle{485} \acmMonth{11}\acmDOI{10.1145/3687024}

\usepackage{siunitx}
\usepackage{makecell}
\usepackage{subcaption}
\usepackage{amsmath}
\usepackage{graphicx}
\usepackage{soul}
\soulregister\cite7
\soulregister\ref7
\soulregister\autoref7

\begin{document}

\title[Beyond Recommendations]{Beyond Recommendations: From Backward to Forward AI Support of Pilots' Decision-Making Process}

\author{Zelun Tony Zhang}
\email{zhang@fortiss.org}
\orcid{0000-0002-4544-7389}
\affiliation{%
  \institution{fortiss GmbH, Research Institute of the Free State of Bavaria}
  \city{Munich}
  \country{Germany}
}
\affiliation{%
  \institution{LMU Munich}
  \city{Munich}
  \country{Germany}
}

\author{Sebastian S. Feger}
\orcid{0000-0002-0287-0945}
\affiliation{%
  \institution{LMU Munich}
  \city{Munich}
  \country{Germany}
}
\affiliation{%
  \institution{TH Rosenheim}
  \city{Rosenheim}
  \country{Germany}}
\email{sebastian.feger@ifi.lmu.de}

\author{Lucas Dullenkopf}
\email{lucas.dullenkopf@airbus.com}
\orcid{0009-0004-0432-1490}
\affiliation{%
  \institution{Airbus Defence and Space GmbH}
  \city{Manching}
  \country{Germany}
}

\author{Rulu Liao}
\authornote{The fourth and fifth authors contributed to this work while working at fortiss GmbH.}
\email{rulu.liao@campus.lmu.de}
\orcid{0009-0005-5521-8730}
\affiliation{%
  \institution{LMU Munich}
  \city{Munich}
  \country{Germany}
}

\author{Lou Süsslin}
\authornotemark[1]
\email{suesslin@pm.me}
\orcid{0009-0004-3922-4578}
\affiliation{%
  \institution{TU Wien}
  \city{Vienna}
  \country{Austria}
}

\author{Yuanting Liu}
\email{liu@fortiss.org}
\orcid{0000-0002-8651-6272}
\affiliation{%
  \institution{fortiss GmbH, Research Institute of the Free State of Bavaria}
  \city{Munich}
  \country{Germany}
}
 
\author{Andreas Butz}
\email{butz@ifi.lmu.de}
\orcid{0000-0002-9007-9888}
\affiliation{%
  \institution{LMU Munich}
  \city{Munich}
  \country{Germany}
}

\renewcommand{\shortauthors}{Zelun Tony Zhang et al.}

\begin{abstract}
    AI is anticipated to enhance human decision-making in high-stakes domains like aviation, but adoption is often hindered by challenges such as inappropriate reliance and poor alignment with users' decision-making. Recent research suggests that a core underlying issue is the recommendation-centric design of many AI systems, i.e., they give end-to-end recommendations and ignore the rest of the decision-making process. Alternative support paradigms are rare, and it remains unclear how the few that do exist compare to recommendation-centric support. In this work, we aimed to empirically compare recommendation-centric support to an alternative paradigm, continuous support, in the context of diversions in aviation. We conducted a mixed-methods study with 32 professional pilots in a realistic setting. To ensure the quality of our study scenarios, we conducted a focus group with four additional pilots prior to the study. We found that continuous support can support pilots' decision-making in a forward direction, allowing them to think more beyond the limits of the system and make faster decisions when combined with recommendations, though the forward support can be disrupted. Participants' statements further suggest a shift in design goal away from providing recommendations, to supporting quick information gathering. Our results show ways to design more helpful and effective AI decision support that goes beyond end-to-end recommendations.
\end{abstract}



\received{January 2024}
\received[revised]{April 2024}
\received[accepted]{May 2024}

\begin{teaserfigure}
    \centering
    \includegraphics[width=0.95\textwidth]{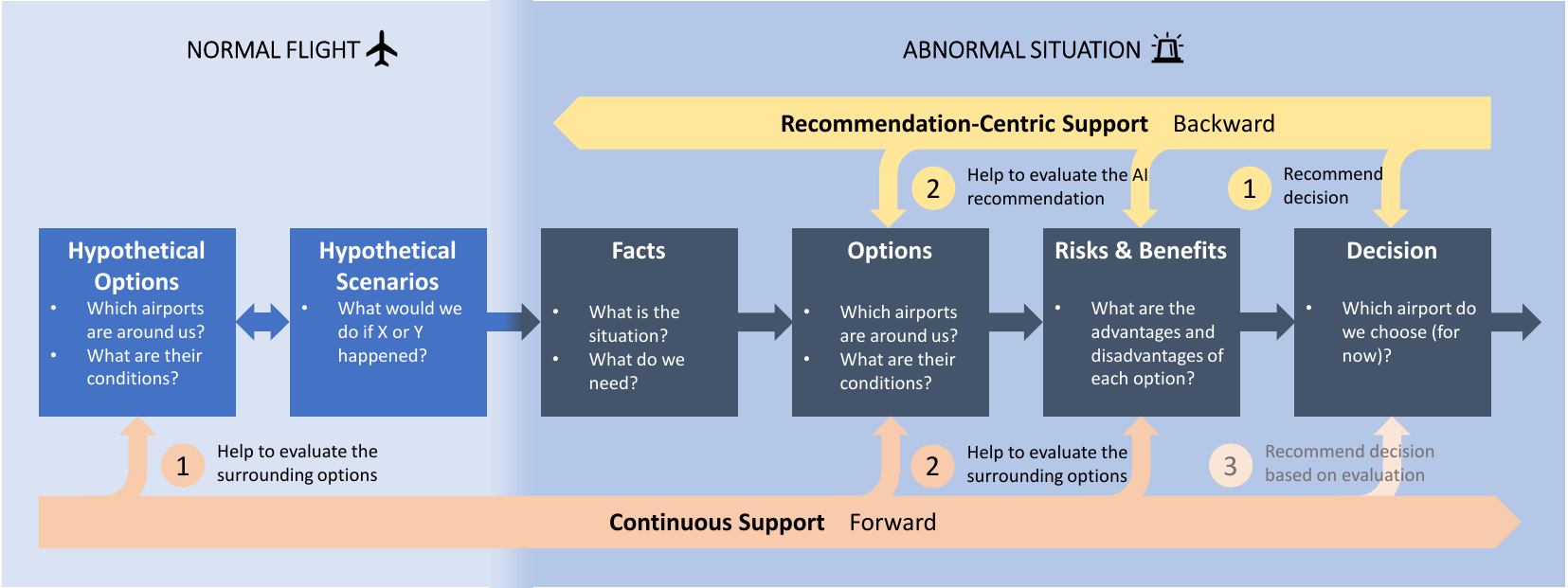}
    \caption{Conceptual overview of the two decision support paradigms that we compare: \textit{recommendation-centric support} and \textit{continuous support}. While the former pushes pilots to reason backward from a decision recommendation, the aim of the latter is to support pilots in a forward direction. Instead of only popping up during an emergency, the system continuously helps pilots to evaluate their surroundings even during normal flight. The system does not give recommendations to avoid biasing pilots. But given continuous support, it may be possible to add recommendations in an emergency while still allowing pilots to reason forward, since pilots are already engaged with the system when they see the recommendations.} 
    \label{fig:concept}
    \Description{Schematic diagram of both support concepts. There two main sections that fade into each other: normal flight to the left, and abnormal situation to the right. In the middle, there are six blocks connected by arrows, which are from left to right: hypothetical options and hypothetical scenarios for normal flight, and facts, options, risks and benefits, and decision for abnormal situation. Above these blocks, there is a thick arrow labeled recommendation-centric support (backward). It points from right to left throughout the abnormal situation section. Smaller arrows split out of it down to the options, risks and benefits, and decision blocks. The one pointing into the decision block is labeled 1) recommend decision. The two other arrows are labeled 2) help to evaluate the AI recommendation. Below the central block is another thick arrow, pointing from left to right throughout the entire figure. One smaller arrow splits out and points into the hypothetical options block, labeled 1) help to evaluate the surrounding options. Two further smaller arrow point into the options and risks and benefits blocks, labeled 2) help to evaluate the surrounding options. A last smaller arrow, which has a slightly faded color, points into the decision block, labeled 3) recommend decision based on evaluation.}
\end{teaserfigure}

\maketitle

\section{Introduction}
AI is projected to improve human decision-making in various high-stakes domains, such as healthcare~\cite{yu_artificial_2018}, finance~\cite{cao_ai_2022}, or law enforcement~\cite{raaijmakers_artificial_2019}. Another domain is aviation~\cite{easa_artificial_2023}, where AI is expected to not only increase decision efficiency, but also safety, as faulty decision-making is one of the main reasons for accidents in aviation~\cite{shappell_human_2007}, motivating research at the intersection between novel technologies and human factors \cite{feger_hci_2022}. One type of decisions in aviation are \textit{diversion} decisions. A diversion is when a flight is unable to reach its planned destination, e.g. due to a technical failure, a medical emergency, or adverse weather conditions. It is the pilots' responsibility to decide on an alternate airport to divert to. While diversions are rare, they are very disruptive and costly for operations~\cite{lewis_data-driven_2021}. Poor diversion decisions can further increase the cost or even impact flight safety. Today, pilots have various tools to support them during diversions, but these are often cumbersome to use and not integrated with each other. Diversions are therefore one primary use case where pilots seek better support and where they can imagine AI assistance~\cite{wurfel_intelligent_2023}.

Yet, in spite of good machine performance, real-world adoption of AI is often difficult~\cite{schlogl_artificial_2019,varghese_artificial_2020}. In controlled studies, failure to achieve \textit{complementary performance} is often observed~\cite{bansal_does_2021,liu_understanding_2021,fok_search_2024}, i.e., the combination of human and AI performs worse than one of them alone. This is the result of \textit{inappropriate reliance}, including both \textit{overreliance} (human relies on AI even when it is disadvantageous to do so)~\cite{bansal_does_2021,liu_understanding_2021,jacobs_how_2021,schmidt_calibrating_2020,bussone_role_2015,bucinca_trust_2021,lai_human_2019,poursabzi-sangdeh_manipulating_2021,wang_are_2021,green_principles_2019} and \textit{underreliance} (human rejects AI even when it would be beneficial to rely on it)~\cite{dietvorst_algorithm_2015,cheng_overcoming_2023,castelo_task-dependent_2019,prahl_understanding_2017}. 

In practice, AI support often turns out less useful to decision makers than imagined~\cite{yang_investigating_2016,blomberg_acting_2018,kawakami_improving_2022}. A growing number of formative studies on real-world tasks with domain experts~\cite{kawakami_improving_2022,kawakami_why_2022,jacobs_designing_2021,sivaraman_ignore_2023,bach_if_2023,burgess_healthcare_2023,kaltenhauser_you_2020,yang_investigating_2016,zhang_rethinking_2024} try to understand what hinders effective use of AI decision support. A frequent problem is that AI decision support is usually designed to be \textit{recommendation-centric}, where the primary functionality of the system is to give end-to-end decision recommendations, i.e., the system suggests a possible end result straight from its input data. By directly jumping to the end result, these systems only support the very end of the decision-making~\cite{bucinca_beyond_2022}, ignoring the entire process leading up to the decision~\cite{yang_investigating_2016,zhang_rethinking_2024,zhang_resilience_2023}.

While the limitations of recommendation-centric support are becoming increasingly more apparent, there is a notable lack of effective solutions to the identified problems. Current research is mostly limited to proposing alternatives for recommendation-centric support on a conceptual level~\cite{zhang_forward_2021,bucinca_beyond_2022,cabitza_need_2021,miller_explainable_2023,koon_human-capabilities_2022}, but concrete examples are few and far between, with most examples stemming from healthcare~\cite{cai_human-centered_2019,van_berkel_designing_2021,yang_unremarkable_2019,lindvall_rapid_2021,gu_augmenting_2023,zhang_rethinking_2024}. Even fewer works evaluate these alternative support paradigms in comparison to recommendation-centric decision support.

One of the few studies exploring alternative decision support roles for AI is Zhang~et~al.'s work on diversion assistance~\cite{zhang_resilience_2023}. Their key insight was that diversion decisions are not a \textit{point}, but a \textit{process} in which pilots take proactive actions. Even during normal flight, pilots constantly establish situation awareness (SA) and prepare a valid plan B, should an emergency occur. Continuously supporting these proactive actions via unobtrusive hints was found to be a promising support role for AI, but it remains unclear how effective it would be in practice, especially compared to the established recommendation-centric paradigm.

In this paper, we sought to empirically compare continuous against recommendation-centric support (see \autoref{fig:concept}) in terms of their effects on pilots' decision-making process, decision outcomes, and decision time in a realistic task setting. We conducted a mixed-methods study with 32 professional pilots, where pilots made a series of diversion decisions with either recommendations, continuous support, a combination of both, or a baseline system with no AI. We aimed for higher ecological validity than in typical AI-assisted decision-making studies. To this end, we validated and refined our scenarios in a focus group with four additional pilots prior to the study.

Our results challenge the common assumption that AI decision support should be recommendation-centric. Continuous support allowed pilots to think more beyond the limits of the system, was better accepted by pilots, and led to faster decisions when combined with recommendations. Our paper makes the following three contributions:
\begin{enumerate}
    \item We add to the rare examples of evaluative studies of AI-assisted decision-making with experts on a real-world task, in a domain that is understudied in the HCI community.
    \item We conduct one of the first empirical comparisons between recommendation-centric and alternative forms of AI decision support, demonstrating the importance and potential of thinking beyond the typical recommendation-centric paradigm.
    \item Based on our results, we propose a framework for \textit{process-oriented decision support} as alternative to recommendation-centric support. We provide continuous support as well as further suggestions from our participants as concrete implementations of process-oriented support for diversions, but we consider the framework to be applicable in other domains beyond aviation as well.
\end{enumerate}

\section{Background and Related Work}
We outline recent work on recommendation-centric decision support in \autoref{sec:recommendation_centric_support} as well as alternatives to this dominant support paradigm in \autoref{sec:alternative_support}. We then describe the gap and the research questions we address in \autoref{sec:research_questions}.



\subsection{Recommendation-Centric Decision Support}
\label{sec:recommendation_centric_support}
The most common strategy to help decision makers work better with recommendation-centric AI is to add explanations of how the model works~\cite{wang_are_2021,bussone_role_2015,bansal_does_2021,zhang_effect_2020,lai_human_2019} and other model information, such as model confidence~\cite{zhang_effect_2020,mcguirl_supporting_2006,rechkemmer_when_2022} or the stated model accuracy~\cite{he_how_2023,rechkemmer_when_2022}. The goal is to help people to rely appropriately on AI recommendations~\cite{schemmer_appropriate_2023} by giving cues about when they may be beneficial or detrimental to rely on. 
Results have been mixed so far, as especially explanations are prone to induce blind trust~\cite{bansal_does_2021,eiband_impact_2019} and hence overreliance, even among domain experts~\cite{bussone_role_2015,jacobs_how_2021}. Recently, Vasconcelos~et~al.~\cite{vasconcelos_explanations_2023} have shown that explanations can reduce overreliance under certain conditions, but it is questionable how often these conditions are valid in real applications~\cite{fok_search_2024}. Communicating model confidence appears more promising, as it has repeatedly been shown to improve appropriate reliance~\cite{zhang_effect_2020,mcguirl_supporting_2006,bansal_does_2021,prabhudesai_understanding_2023}; but this relies on well-calibrated confidence scores, which are often difficult to achieve---models can be wrong with high confidence.

The limited success of adding model information appears to be due to people not engaging cognitively with it~\cite{bucinca_trust_2021,gajos_people_2022}. One way to increase engagement is to employ \textit{cognitive forcing} interventions, such as showing recommendations only after users made an initial decision~\cite{bucinca_trust_2021,fogliato_who_2022}, introducing a waiting time until recommendations are shown~\cite{bucinca_trust_2021,park_slow_2019}, or forcing users to wait before they can proceed to the next task~\cite{rastogi_deciding_2022}. While effective, these interventions negatively impact user experience~\cite{bucinca_trust_2021,fogliato_who_2022,park_slow_2019}. Liu~et~al~\cite{liu_understanding_2021} explored the use of interactive explanations to increase engagement, though it did not reduce overreliance in their case.

Recently, an increasing number of voices call the entire premise of recommendation-centric support into question. Koon~\cite{koon_human-capabilities_2022} emphasizes that decisions are often complex, and that condensing them into an AI recommendation is necessarily reductionist. At the same time, since recommendations are hard to appropriate, users often struggle to combine them with the wider context knowledge they have~\cite{cabitza_need_2021}. From a cognitive science perspective, Miller~\cite{miller_explainable_2023} argues that recommendation-centric support does not align with the cognitive processes of human decision-making. Instead, recommendations take control away from decision makers. Similarly, Wang~et~al.~\cite{wang_designing_2019} and Zhang~et~al.~\cite{zhang_forward_2021} caution against error-prone \textit{backward reasoning} from the end result back to the input data, which is facilitated by a fixation on end-to-end recommendations. All of these authors call for alternative approaches to AI decision support that are less centered on recommendations.

\subsection{Alternative Forms of AI Decision Support}
\label{sec:alternative_support}
One stream of work that de-emphasizes recommendations rethinks the purpose of explanations. Instead of explaining the AI model, explanations can provide information that is of natural interest in a decision, such as domain-specific information~\cite{yang_harnessing_2023,lim_diagrammatization_2023}, or the socio-organizational context of a decision~\cite{ehsan_expanding_2021}. These explanations situate recommendations within the primary decision-making task, rather than diverting attention to the secondary task of understanding the AI model.

Other authors propose entirely different roles for AI than providing end-to-end recommendations. Alternative AI support paradigms are far from new~\cite{woods_paradigms_1986}, but have been largely ignored by current research. This is arguably more due to technical feasibility rather than consideration for human needs, since with modern AI methods, it is straightforward to formulate many decision tasks as end-to-end predictions~\cite{bucinca_beyond_2022,shneiderman_design_2020}. As for alternatives to providing recommendations, on a conceptual level, Cabitza~et~al.~\cite{cabitza_need_2021} propose to frame AI as ``knowledge artifact functions'' which support people in their collaborative decision-making. Zhang~et~al.~\cite{zhang_forward_2021} put forward what they call ``forward-reasoning decision support'', where users form decisions themselves, augmented by rich interactions with AI tools. In a similar, but more concrete way, Miller~\cite{miller_explainable_2023} proposes the concept of ``evaluative AI'', which focuses on helping decision makers to evaluate different hypotheses. 

In essence, all of these proposals aim to help people to make decisions through \textit{forward reasoning}, allowing people to start from the context at hand and to use their domain expertise to reach a decision. This is in contrast to recommendation-centric support, which pushes people to reason backward from the recommendation. In fact, cognitive forcing can be seen as a way to encourage forward reasoning as well by pushing people to think independently from AI recommendations. However, with cognitive forcing interventions, people get no support while making their independent decisions, which reduces the supportive value of the AI~\cite{bucinca_trust_2021,park_slow_2019,fogliato_who_2022}. The above concepts aim to facilitate forward reasoning while supporting users' decision-making processes.

Beyond abstract concepts, concrete examples for supporting decisions without relying on end-to-end recommendations are often found in healthcare. Lindvall~et~al.~\cite{lindvall_rapid_2021} designed a system for tumor assessment that navigates pathologists to potentially tumorous image regions for them to review, without revealing whether or how confidently the model classifies the pixels as tumor. Crucially, the classification threshold was not set to optimize accuracy, but sensitivity. Consequently, if pathologists did not find a tumor in any of the suggested regions, they could be relatively sure that the rest of the image also contains no cancer. Zhang~et~al.~\cite{zhang_rethinking_2024} studied the case of sepsis diagnosis, where an existing AI tool only addresses the final stage of the decision-making by offering a sepsis risk score and sending alerts above a certain threshold. The authors proposed a redesign where the system suggests lab tests that would help to reduce uncertainty about patients' future conditions.

In the context of aviation, Zhang~et~al. explored how to support diversions decisions~\cite{zhang_resilience_2023}. 
Their system does not recommend airports, but continuously provides unobtrusive local hints about potential limitations at the surrounding airports. The system provides this support of pilots' SA also during normal flight when there is no sign of an emergency yet.

\subsection{Summary and Research Questions}
\label{sec:research_questions}
There is a remarkable difference between the studies mentioned in the previous two sections: The studies on recommendation-centric support in \autoref{sec:recommendation_centric_support} were almost all built on simple, artificial tasks with lay persons as participants. Such studies make up the majority of research in AI-assisted decision-making~\cite{lai_towards_2023}. In contrast, the alternative support paradigms in \autoref{sec:alternative_support} all stemmed from studying complex real-world decisions with domain experts. This indicates a potentially significant gap between what is studied in large-scale controlled experiments and what is actually required by experts in real applications. Our study is situated right in this gap by conducting a controlled comparison between alternative decision support paradigms with domain experts.

The goal of our work is to empirically compare continuous support with typical recommendation-centric support. We further add a combination of both to the comparison, where the system provides continuous support during normal flight, but gives recommendations in an emergency. \autoref{fig:concept} contrasts how the two paradigms conceptually fit into pilots' decision-making, based on the \textit{FOR-DEC} model~\cite{hormann_for-dec_1994}. FOR-DEC is a prescriptive model used by many airlines to train pilots to make decisions in a structured manner. It is an acronym for the steps to follow during decision-making: \textit{facts}, \textit{options}, \textit{risks \& benefits}, \textit{decision}, \textit{execution}, \textit{check}. The dash in the middle signals a pause for reflection before making the decision. \autoref{fig:concept} only covers the first four steps of the model, since execution and check are beyond the scope of our work. 

The focus of our empirical comparison is to assess whether pilots reason differently with continuous support and recommendation-centric support, and how this affects pilots' perceptions about the system as well as decision outcomes, namely overreliance and decision time. As outlined in \autoref{sec:recommendation_centric_support}, the former is one of the major concerns in recommendation-centric support. The latter is of interest since aviation is highly dynamic, which limits the time available for decision-making. We therefore pose the following research questions:
\begin{itemize}
    \item \textbf{RQ1}: How do pilots integrate the different support paradigms into their workflow?
    \item \textbf{RQ2}: How do the support paradigms differ in terms of overreliance?
    \item \textbf{RQ3}: How do the support paradigms differ in terms of decision time?
    \item \textbf{RQ4}: How do pilots perceive the different support paradigms?
\end{itemize}
We hypothesize that recommendations induce backward reasoning, while continuous support facilitates forward reasoning. We further hypothesize that combining both allows to give recommendations while still enabling forward reasoning. Hence, we expect continuous support to lead to less overreliance than recommendations, but at the expense of longer decision times. We expect the combination of both to combine their advantages, i.e., less overreliance and shorter decision times.

Most closely related to our work is a study by Smith~et~al.~\cite{smith_brittleness_1997}, who compared the effects of three different versions of a flight planning tool on users' performance. Apart from being a different task, the major difference to our work is that their system versions represent consecutively increasing levels of automation, while we study alternative support paradigms.

\section{Diversion Assistance System}
In this section, we present the design of the four versions our diversion assistance system (\autoref{sec:system_design}), followed by the apparatus used to conduct our study (\autoref{sec:apparatus}).

\subsection{System Design and Variants}
\label{sec:system_design}
Our diversion assistance system (DAS) consists of three basic components (\raisebox{.5pt}{\textcircled{\raisebox{-.9pt} {6}}} in \autoref{fig:screenshots}). A navigation map in the top left of the main screen shows surrounding airports relative to the current aircraft position. The bottom half of the main screen shows the same airports in a table, displaying runway, weather, and operational information as well as the time to and fuel remaining at each airport. The table can be sorted by any of the columns by tapping the respective column header. The details box to the right of the navigation map shows additional raw information for the selected airport, including raw runway data, weather reports, and NOTAMs\footnote{\textit{Notice to Air Missions}, formerly \textit{Notice to Airmen}. These are messages containing information about abnormal conditions that may affect flight operations, e.g. runway closures, inoperable systems at an airport, or drones near the runway.}.

\begin{figure*}[thb]
    \includegraphics[width=\textwidth]{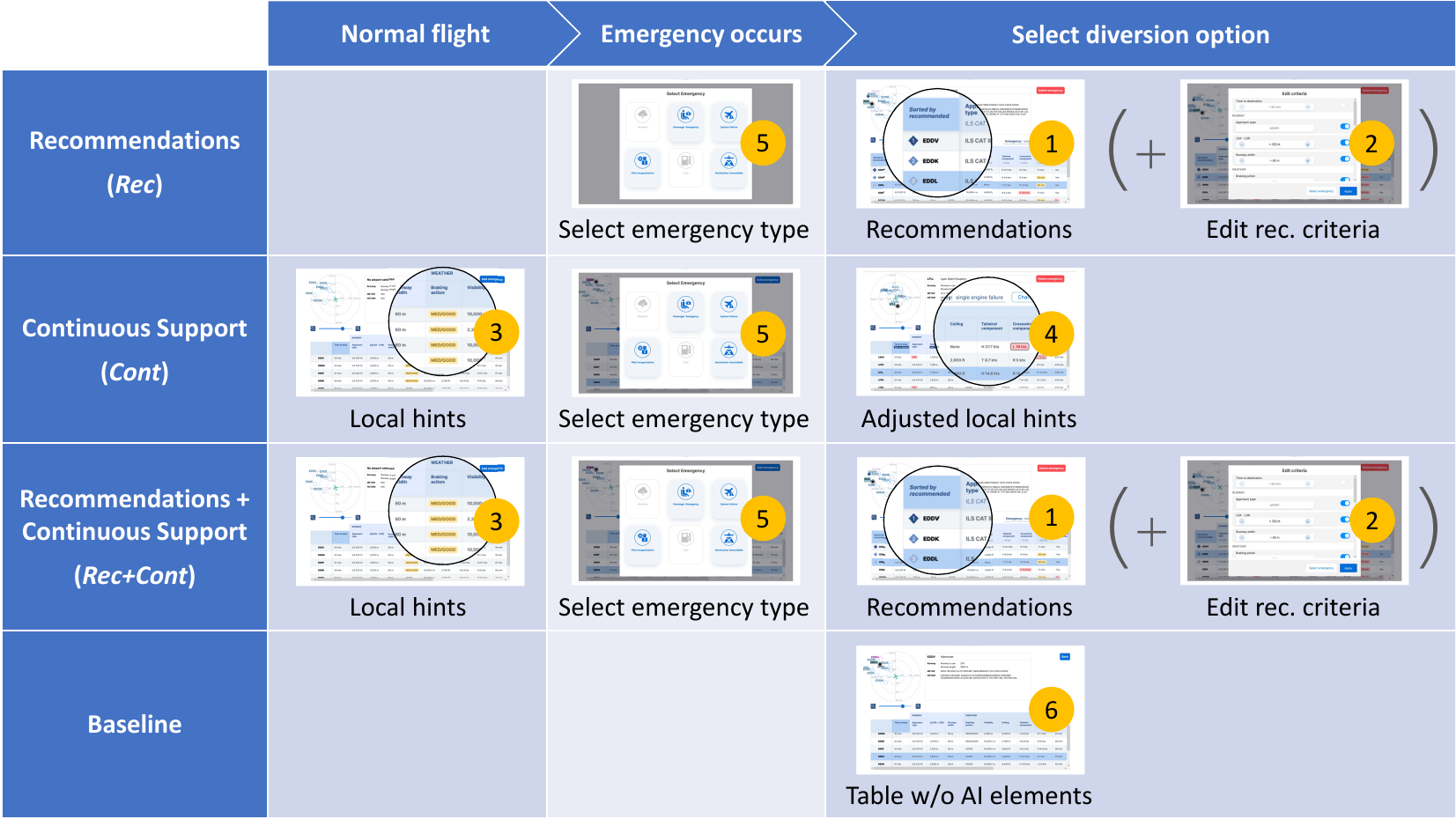}
    \caption{The four versions of the diversion assistance system, from top to bottom: \textit{Rec}, \textit{Cont}, \textit{Rec+Cont}, \textit{Baseline}. The corresponding views \raisebox{.5pt}{\textcircled{\raisebox{-.9pt} {1}}}--\raisebox{.5pt}{\textcircled{\raisebox{-.9pt} {6}}} that are applicable during the three operational phases \textit{normal flight}, \textit{emergency occurs}, and \textit{select diversion option}, are numbered corresponding to the detailed view in \autoref{fig:screenshots}.}
    \label{fig:conditions}
    \Description{Table with four versions of the diversion assistance. The rows are the four versions, the columns are normal flight, emergency occurs, and select diversion option. The table shows smaller versions of the screeshots shown in Figure 3. The normal flight column is populated with screenshots of the local hints for Cont and Rec+Cont, and empty for the other two rows. The emergency occurs column holds screenshots for select emergency type for all rows except for Baseline. The select diversion option column holds for the Rec and Rec+Cont rows screenshots for recommendations and in brackets for edit recommendation criteria. For the Cont row, there is screenshot for adjusted local hints. For the Baseline row, there is a screenshot for table without AI elements.}
\end{figure*}
\begin{figure*}[thb]
    \includegraphics[width=\textwidth]{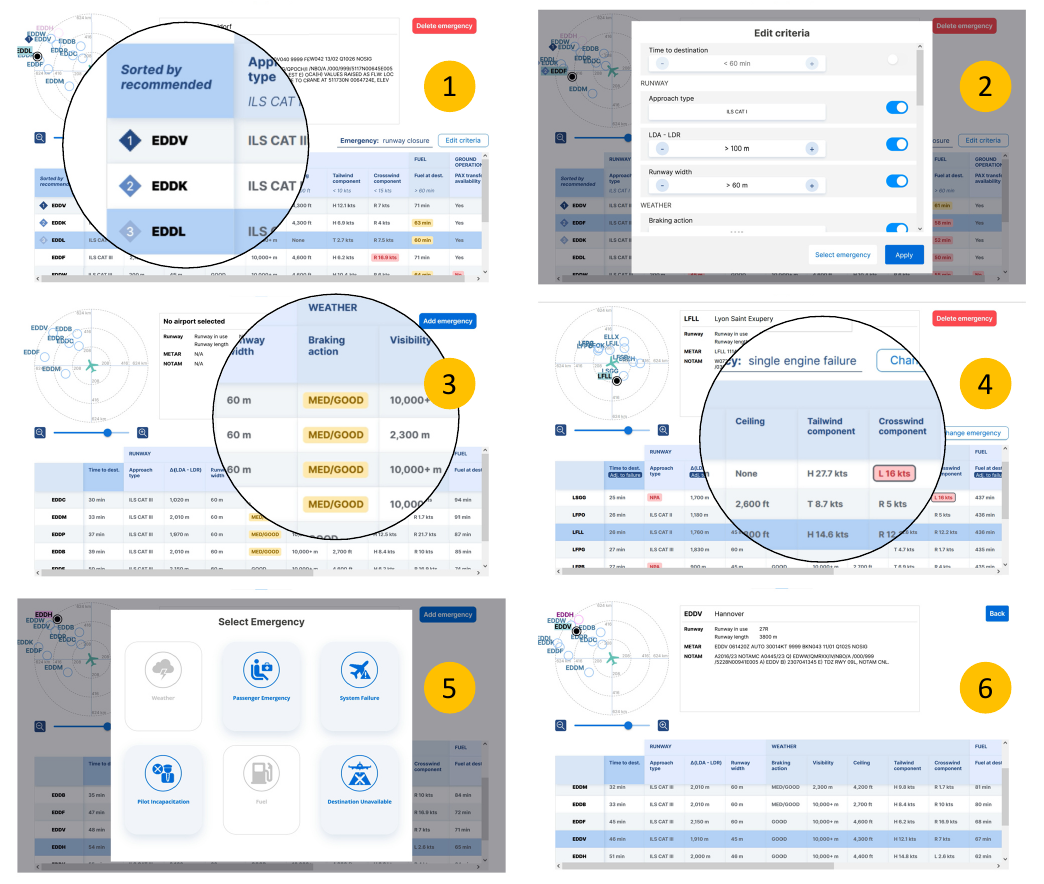}
    \caption{Screenshots of the diversion assistance system. \raisebox{.5pt}{\textcircled{\raisebox{-.9pt} {1}}} Recommendations. \raisebox{.5pt}{\textcircled{\raisebox{-.9pt} {2}}} Edit recommendation criteria. \raisebox{.5pt}{\textcircled{\raisebox{-.9pt} {3}}} Local hints during normal flight. \raisebox{.5pt}{\textcircled{\raisebox{-.9pt} {4}}} Adjusted local hints during emergency. \raisebox{.5pt}{\textcircled{\raisebox{-.9pt} {5}}} Select emergency type. \raisebox{.5pt}{\textcircled{\raisebox{-.9pt} {6}}} Baseline without AI. The numbers correspond to those in \autoref{fig:conditions}.}
    \label{fig:screenshots}
    \Description{Six numbered screenshots of the system, matching the numbers in the caption, arranged in a grid with two columns and three rows. Screenshots 1 (recommendations), 3 (local hints during normal flight), and 4 (adjusted local hints during emergency), magnify in circles the respective parts of the screenshots.}
\end{figure*}
Based on these basic components, we designed four versions of the DAS, as shown in \autoref{fig:conditions}. In the \textbf{Recommendations} version, the system sorts the table by the AI's evaluation of each airport and recommends up to three options, which are marked in both the table and the navigation map (\raisebox{.5pt}{\textcircled{\raisebox{-.9pt} {1}}} in \autoref{fig:screenshots}). The AI's evaluations are based on pre-defined criteria, like having a certain amount of fuel left at an airport. These criteria are shown in their respective column header and can be edited by pilots if desired (\raisebox{.5pt}{\textcircled{\raisebox{-.9pt} {2}}} in \autoref{fig:screenshots}). Yellow and red highlights indicate when a criterion is only barely met or not met at an airport, providing transparency about the AI's evaluation. The table can be sorted manually by each criterion by tapping the respective column header. The intended flow of this system version is as follows: When an emergency or abnormal situation happens, pilots first select the emergency type (\raisebox{.5pt}{\textcircled{\raisebox{-.9pt} {5}}} in \autoref{fig:screenshots}), after which the main screen with the recommendations shows up. The pre-set evaluation criteria are catered to the selected emergency type. This system variant represents the common recommendation-centric paradigm of AI decision support.

In contrast, the \textbf{Continuous Support} version shows the main screen also in normal flight when there is no emergency in sight yet. In this normal flight mode, the system continuously evaluates the surrounding airports for potential constraints that are of general interest for any type of diversion, like a wet runway (\raisebox{.5pt}{\textcircled{\raisebox{-.9pt} {3}}} in \autoref{fig:screenshots}). These potential constraints are shown in the table, again as yellow and red highlights. Instead of explaining recommendations---which this system variant does not have---the highlights serve as local hints to guide pilots' attention. This normal flight mode is meant to support pilots' SA and their continued preparation for hypothetical emergencies. 

In an emergency, pilots can switch to emergency mode, again by selecting the emergency type, which adjusts the hints to the selected emergency (\raisebox{.5pt}{\textcircled{\raisebox{-.9pt} {4}}} in \autoref{fig:screenshots}). For instance, 16 knots of crosswind may not be a big concern normally, but with an engine failure, it might be critical. The system would highlight the crosswind, which it did not highlight in normal flight. New hints are indicated by a solid black border. At no point does this system variant generate decision recommendations. The table is always sorted by the pilot-selected column, and per default by \textit{time to destination}.

The \textbf{Recommendations + Continuous Support} version combines the two prior versions,
with continuous support in normal flight, and recommendations in case of an emergency. Lastly, we added a \textbf{Baseline} version, which only shows the main screen with neither recommendations nor local hints (\raisebox{.5pt}{\textcircled{\raisebox{-.9pt} {6}}} in \autoref{fig:screenshots}) and which is only available in case of emergency.

\subsection{Apparatus}
\label{sec:apparatus}
\begin{figure*}[thb]
    \includegraphics[width=0.9\textwidth]{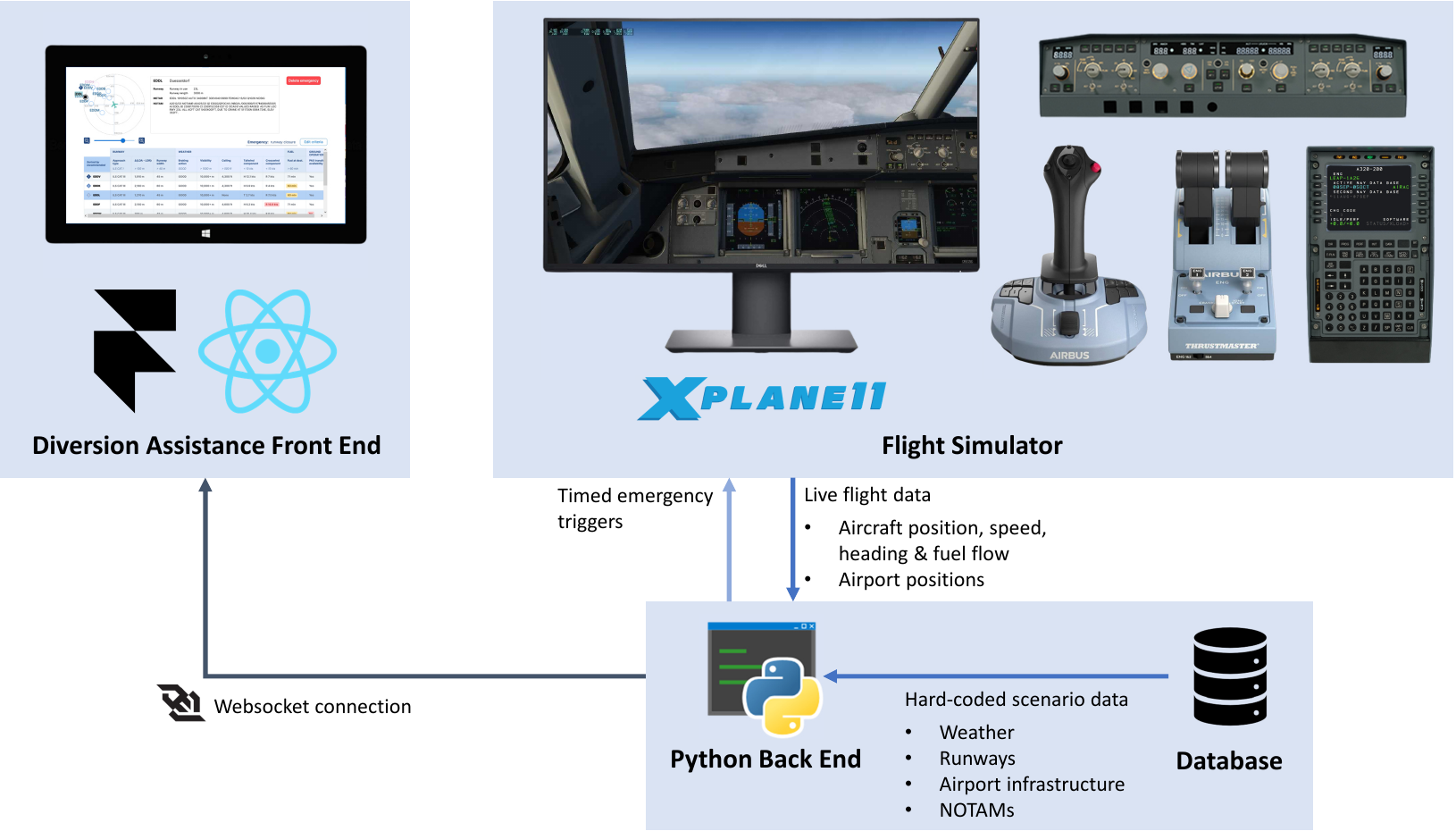}
    \caption{Overview of the study apparatus.}
    \label{fig:setup}
    \Description{Three main blocks. Top left: diversion assistance front end, with a tablet and logos of Framer and React. Top right: flight simulator, with a monitor showing the A320 cockpit in X-Plane, the X-Plane 11 logo, and depictions of the FCU, the MCDU, throttle, and sidestick. At the bottom: block with database and Python back end, with arrow pointing from the former to the latter. The arrow is labeled hard-coded scenario data, with bullets reading weather, runways, airport infrastructure, NOTAMs. An arrow points from the bottom block to the front end block, labeled Websocket connection. Another arrow points from the bottom block to the flight simulator block, labeled timed emergency triggers. Another arrow points the opposite direction, labeled live fight data, with bullets reading aircraft position, speed, heading and fuel flow; airport positions.}
\end{figure*}

As simulation environment, we chose X-Plane 11\footnote{https://www.x-plane.com/}, running the Airbus A320 Ultimate aircraft model by Flight Factor\footnote{https://flightfactor.aero/}. We supplemented the simulator with hardware controls, including a sidestick and a throttle quadrant from Thrustmaster\footnote{https://www.thrustmaster.com/}, as well as an MCDU\footnote{\textit{Multipurpose Control and Display Unit}, the input device for the flight management system and other computer systems.} and an FCU\footnote{\textit{Flight Control Unit}, the control panel used to control the autopilot.} from Skalarki\footnote{https://www.skalarki-electronics.com/}, to reduce reliance on a mouse for interacting with the aircraft. The DAS was implemented as an interactive mockup in the prototyping tool Framer\footnote{https://www.framer.com/}, with custom components written in React. It runs on a second-generation Surface Pro tablet next to the simulator computer, mimicking an EFB\footnote{\textit{Electronic Flight Bag}, a portable device to store and display flight-relevant data and to run assistive applications.}. The mockup communicates via Websocket with a Python back end on the simulator computer, which acts as an interface to X-Plane to read out live flight data. Weather and ground facilities at each airport are hard-coded for each scenario. The Python back end further triggers the emergencies of the study scenarios in X-Plane based on timers. The entire setup is shown in \autoref{fig:setup}. Note that despite the remarkable realism of the elements we used in our simulation setup, some notable deviations from real flights remained. We describe these deviations and discuss their implications in \autoref{sec:limitations}.

The AI recommendations are generated with a manually-tuned scoring function that assigns each airport a score, calculated as a weighted sum of subscores for each criterion. Each subscore is calculated using a piecewise linear function that penalizes unfulfilled criteria more heavily than it rewards overfulfilled criteria. Using this scoring function, the recommendations react dynamically to what is happening in the simulator and to the criteria the pilot defines. While it is not machine learning, given the large number of criteria, it is still hard to comprehend what exactly causes a certain airport to be recommended or not, just as with an opaque deep neural network.

\section{Methods}
\begin{figure*}[thb]
    \includegraphics[width=0.9\textwidth]{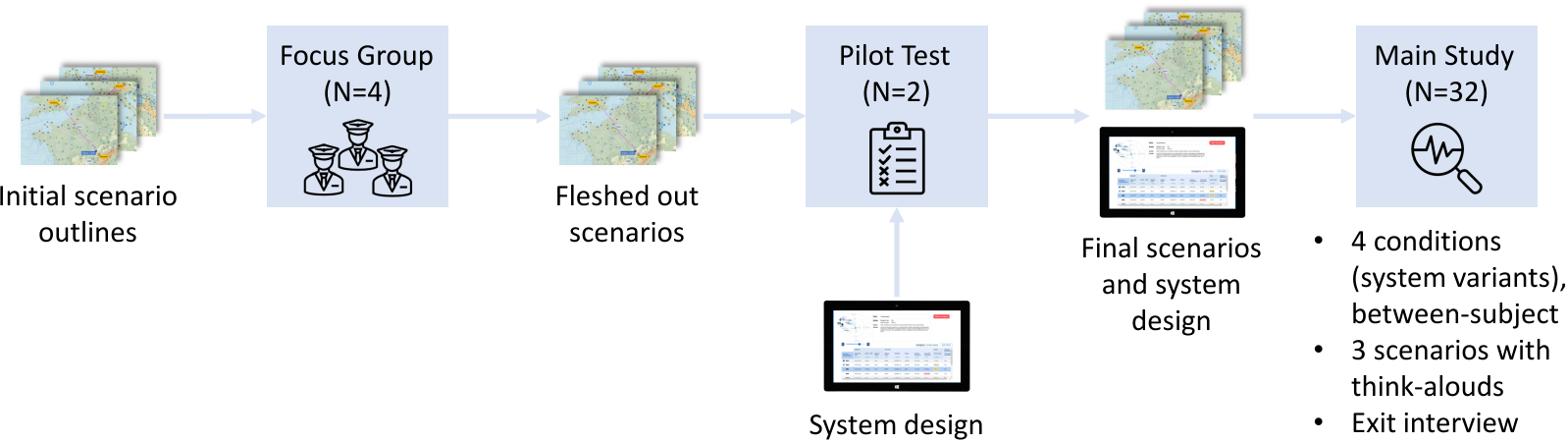}
    \caption{Overview of the study parts.}
    \label{fig:study_parts}
    \Description{Pipeline starting with initial scenario outlines on the left, followed by a right-pointing arrow, followed by focus group (N=4), followed by another right-pointing arrow, followed by pilot test (N=2). Below pilot test, there is a block for system design, with an arrow pointing up to pilot test. To the right of pilot test is another right-pointing arrow, followed by final scenario and system design, followed by yet another right-pointing arrow and main study (N=32). Below main study are three bullet points, reading 4 conditions (system variants), between-subject; 3 scenarios with think-alouds; exit interview.}
\end{figure*}

Besides the DAS, valid and informative scenarios were the key component of our study. To ensure the quality of our scenarios, we first conducted a focus group to discuss and flesh out our initial scenario outlines with professional pilots. We subsequently tested our scenarios and system design with pilots before finally running our main study. In this section, we describe these steps and our methodology in detail, which are shown in \autoref{fig:study_parts}.

\subsection{Focus Group and Study Scenarios}
\label{sec:focus_group}
There is no inherent right or wrong in most diversion decisions, so one goal of the focus group was to obtain a reference for how pilots would decide in each scenario and for which reasons. The other goal was to understand how to design the details of the scenarios to fit our intentions. We recruited four professional pilots for the focus group (one captain, three first officers; all male; median age: 29.5 years (IQR 28.5--34.75); median flight hours: 1000 hours (IQR 392.5--3750); details in \autoref{sec:participant_details}, \autoref{tab:focus_participants}). All four work at the same airline, but have past experience in four additional airlines. We showed them the initial outline of each scenario and asked them to discuss how they would decide. The focus group had a duration of 90 minutes, and participants were compensated with 150~EUR ($\approx$ 164~USD) each, which is a typical rate for professional pilots, given the difficulty to recruit them.

After the focus group, we adjusted the scenario outlines and filled in the details according to the insights gained. In total, we designed three scenarios, with the aim to cover a good range of common reasons for diversions as well as different potential failure modes of the DAS, as recommended by Roth~et~al.~\cite{roth_methods_2022}. In particular, we aimed to construct one scenario each where the DAS performs well (Scenario 1), suggests an airport that pilots tend to disagree with (Scenario 2), and suggests a solution that is subpar, but for a reason that is not immediately obvious (Scenario 3).

\subsubsection{Scenario 1: Single engine failure}
\label{sec:scenario_1}
\begin{figure*}[thb]
    \includegraphics[width=0.6\textwidth]{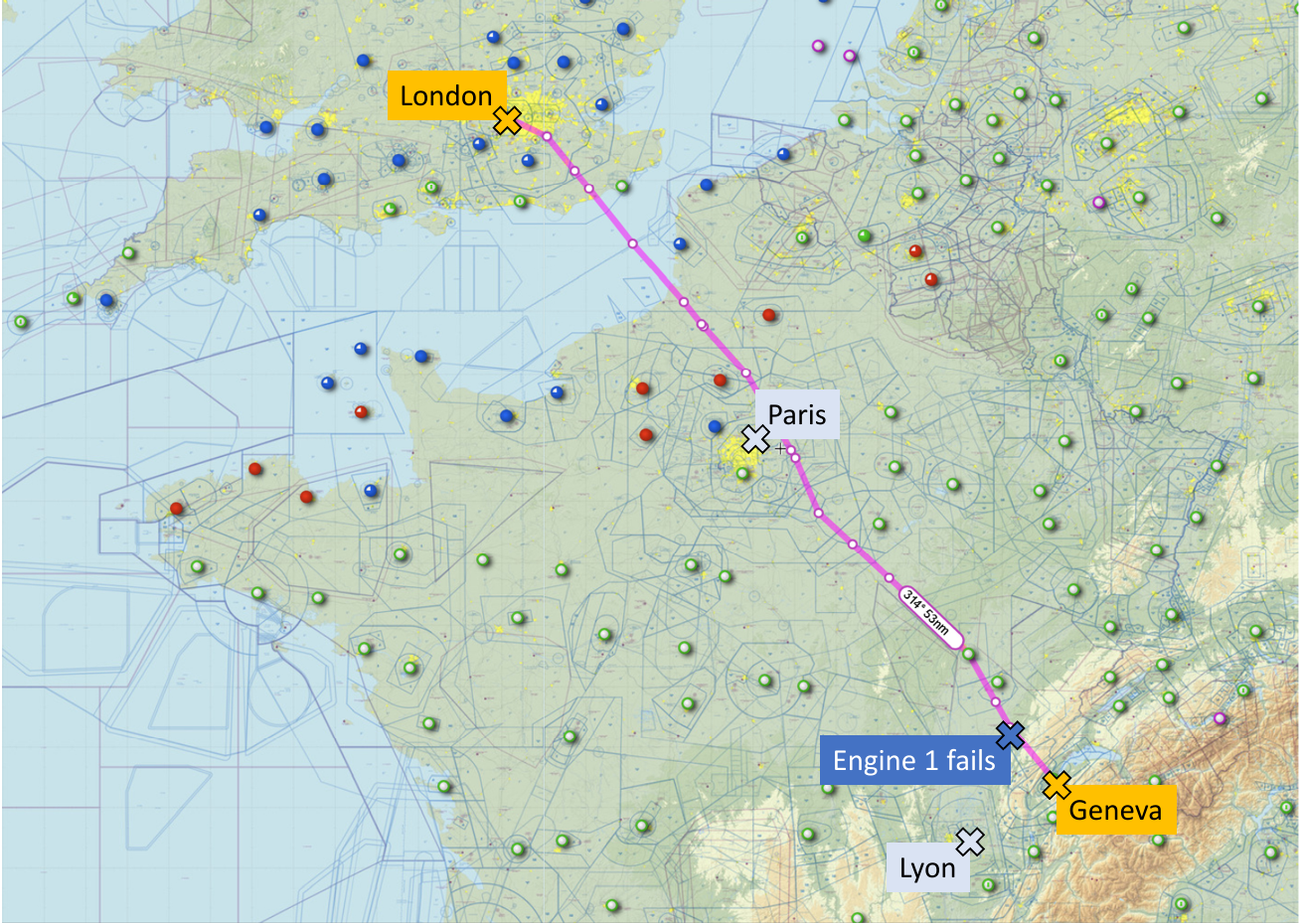}
    \caption[Single engine failure scenario. The flight is from Geneva to London. Image based on SkyVector.]{Single engine failure scenario. The flight is from Geneva to London. Figure based on SkyVector\footnotemark.}
    \label{fig:scenario_1}
    \Description{Screenshot of single engine failure scenario, showing on a map with a line the flight route, and with colored crosses departure (Geneva) and destination (London) as well as notable en-route airports (Paris, Lyon), and the place where the event (Engine 1 fails) happened.}
\end{figure*}
\footnotetext{https://skyvector.com/}In the first scenario, a flight from Geneva, Switzerland to London Heathrow suffers a single engine failure during climb (\autoref{fig:scenario_1}). The scenario is based on incident reports found with the CAROL query tool of the National Transportation Safety Board\footnote{https://carol.ntsb.gov/}. The focus group participants pointed out that the terrain around Geneva may be challenging with an engine failure, but returning to Geneva is likely preferable for the company. Charles de Gaulle on the other hand is a good option since it is a major hub with excellent infrastructure.

Our intention for this scenario was that the DAS would recommend Charles de Gaulle as an option that pilots would agree with. Beginning with this scenario was meant to establish pilots' initial trust into the system, since in reality, such a system would perform well in most situations. To make the decision less ambiguous and hence the recommendation more acceptable, we added a moderately high crosswind component to Geneva. In the scenario, we triggered an engine failure and displayed a popup message within X-Plane, reading \textit{``Engine 1 has failed. No engine relight\footnote{In reality, pilots would try to relight the engine. We asked them to skip this to simplify the scenario, as it would not have added any value to our study.}. Please secure the engine and proceed with the diversion decision.''} Pilots were asked to handle the engine failure within X-Plane as they would in reality to keep them in their known workflow.

\subsubsection{Scenario 2: Passenger medical emergency}
\label{sec:scenario_2}
\begin{figure*}[thb]
    \includegraphics[width=0.6\textwidth]{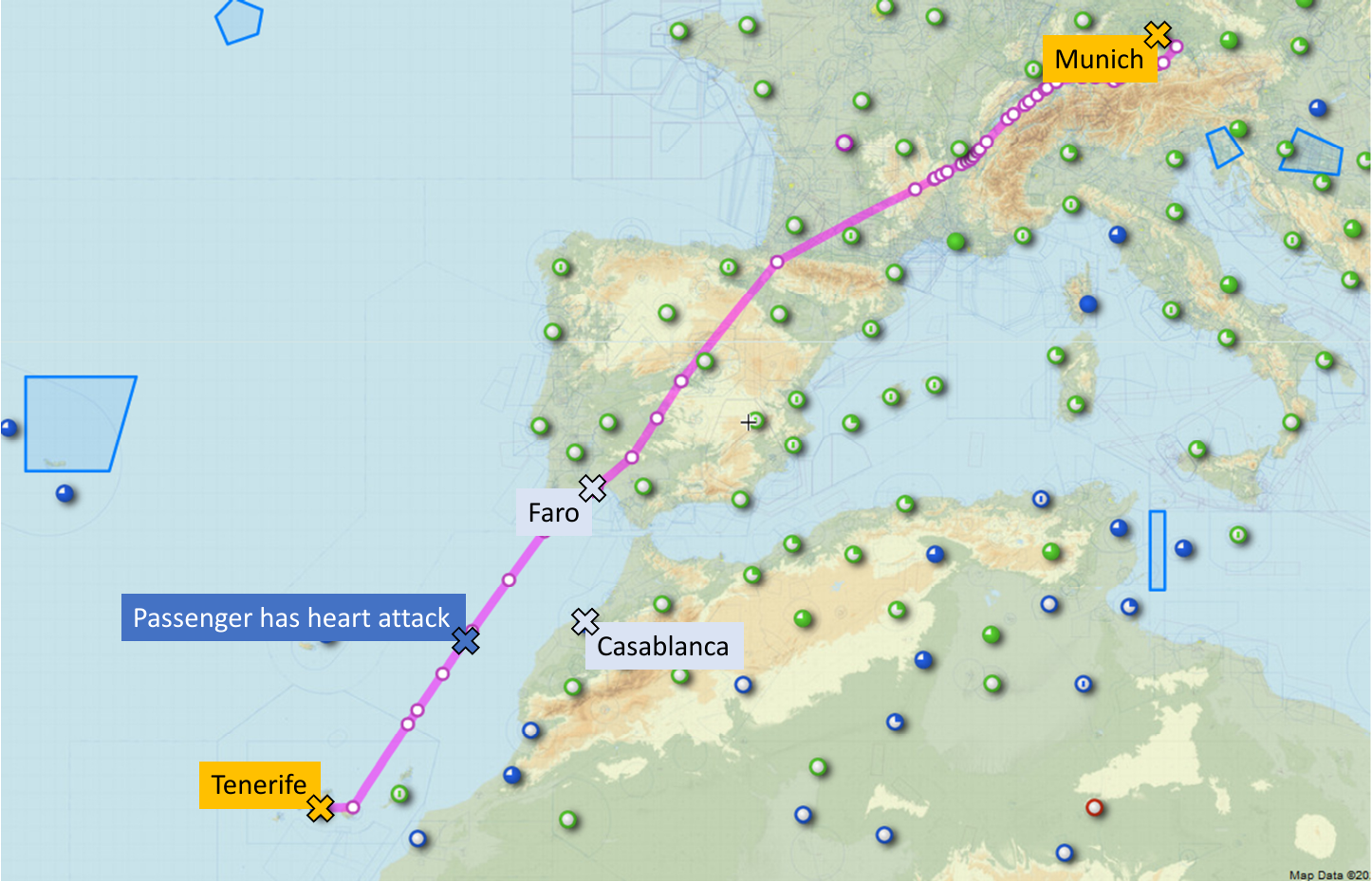}
    \caption{Medical emergency scenario. The flight is from Tenerife to Munich. Figure based on SkyVector.}
    \label{fig:scenario_2}
    \Description{Screenshot of passenger medical emergency scenario, showing on a map with a line the flight route, and with colored crosses departure (Tenerife) and destination (Munich) as well as notable en-route airports (Casablanca, Faro), and the place where the event (passenger has heart attack) happened.}
\end{figure*}
The second scenario was a flight from Tenerife, Spain to Munich, Germany, with a passenger having a heart attack while the flight is above the sea in the middle between Tenerife and the European continent (\autoref{fig:scenario_2}). The emergency was announced through a popup message in X-Plane: \textit{``The passenger on 11C has vomited, complains of extreme chest pains, is pale and sweating. This has been going on for a few minutes now. A doctor (general practitioner) sitting next to the passenger reacted immediately. He suspects a heart attack.''} This scenario was based on a route that is notorious among short-haul flight pilots due to the long flight over open sea, with much fewer diversion options than usual on European short-haul flights.

Our expectation was that Casablanca, Morocco would be the obvious option since it is the closest, making the decision too easy. Hence, we thought about adding a reason to make Casablanca unattractive, like a nation-wide healthcare strike. However, our focus group participants said they would either fly to Faro, Portugal, or return to Tenerife, depending on which is faster, since they would prefer to stay within Europe. This would make it easier to continue the flight and was also assumed to allow for better medical treatment for the passenger. We therefore decided not to add any factor to make Casablanca unattractive. Still, the DAS would recommend Casablanca to probe for pilots' reactions when the system recommends an option that they are not comfortable with.

\subsubsection{Scenario 3: Airport closure}
\label{sec:scenario_3}
\begin{figure*}[thb]
    \includegraphics[width=0.6\textwidth]{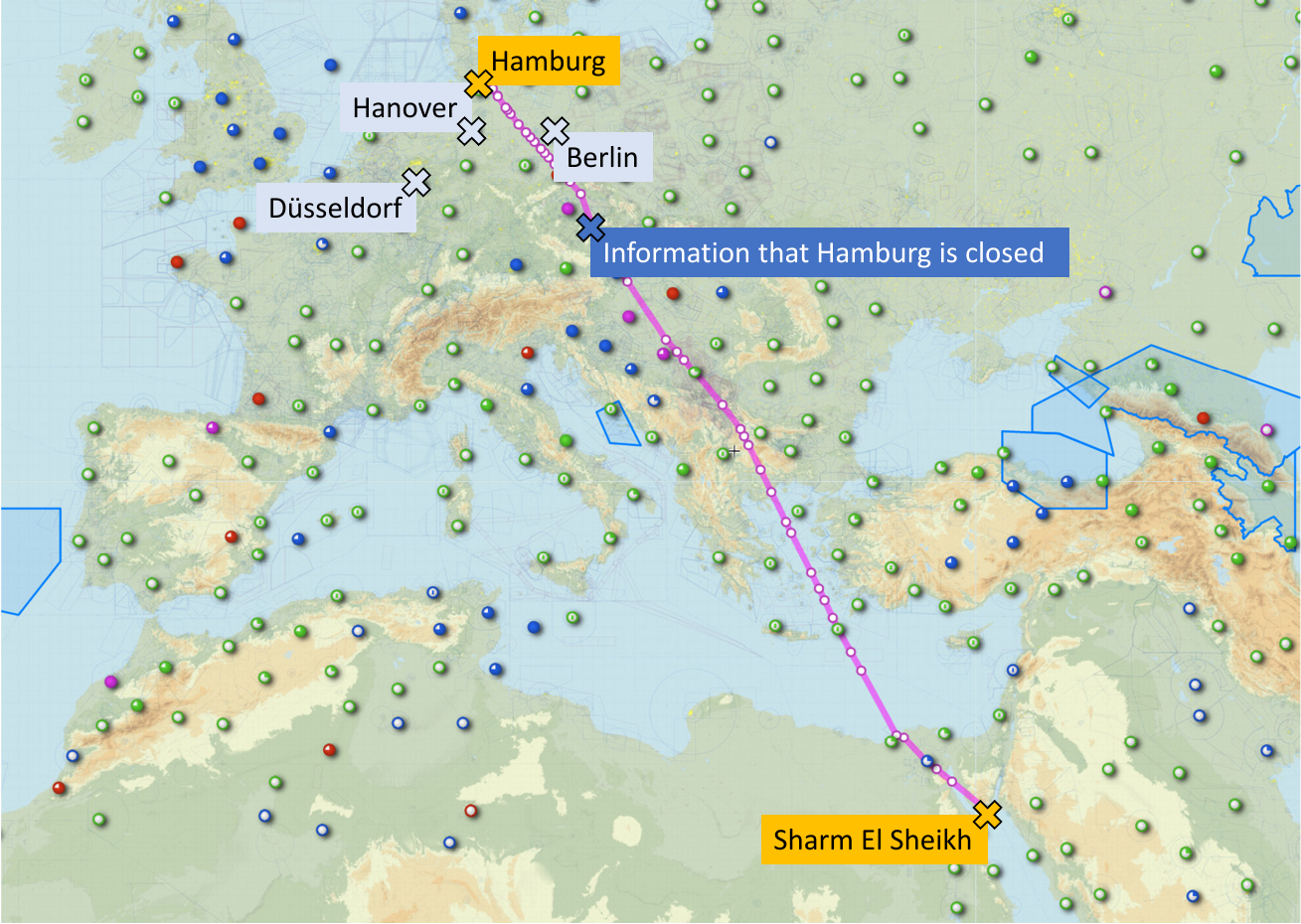}
    \caption{Airport closure scenario. The flight is from Sharm El Sheikh to Hamburg. Figure based on SkyVector.}
    \label{fig:scenario_3}
    \Description{Screenshot of airport closure scenario, showing on a map with a line the flight route, and with colored crosses departure (Sharm El Sheikh) and destination (Hamburg) as well as notable en-route airports (Berlin, Hanover, Düsseldorf), and the place where the event (information that Hamburg is closed) happened.}
\end{figure*}
This last scenario was based on the personal experience of one participant in Zhang~et~al.'s study~\cite{zhang_resilience_2023}. The flight was from Sharm El Sheikh, Egypt, to Hamburg, Germany (\autoref{fig:scenario_3}). About an hour before landing, participants were given the information that the destination had closed down, again via a popup message: \textit{``Due to a power outage, EDDH\footnote{The ICAO (International Civil Aviation Organization) airport code for Hamburg.} had to close completely. It will remain closed for at least the next few hours.''}

The intention for this scenario was that the DAS would recommend an option that is good at first glance, but problematic at second thought, in order to probe for how much participants would over-rely. The system would recommend Hanover, Germany, which is the typical alternate for Hamburg. It also has good ground connections to Hamburg for the passengers and is therefore a plausible recommendation. However, the airport closure affects the entire air traffic coming into Hamburg. Many planes would try to divert to Hanover, which in the real event had caused Hanover to run full. The DAS does not consider this traffic factor.

Unprompted, focus group participants discussed the traffic situation in Hanover as an important consideration, confirming that pilots would realistically think about it in such a situation. Participants said that instead of Hanover, all the en-route airports like Berlin, Dresden, or Leipzig would be valid options. We added slightly unfavorable, but workable weather conditions at these airports, to make the system recommendation Hanover seem more attractive at first glance.

\subsection{Main Study}
The main study had a between-subject design where each participant was assigned to one of the four DAS variants in \autoref{fig:conditions} and completed all three scenarios in the order of \autoref{sec:focus_group}. 
At the beginning of the study, pilots got a detailed introduction into X-Plane and the DAS. Before starting each scenario, pilots were shown maps similar to those in \autoref{fig:scenario_1}--\autoref{fig:scenario_3}, with the flight plan, the departure and destination airport, as well as the position of the aircraft at the beginning of the scenario. The scenarios started a couple minutes before the incident to allow participants to familiarize themselves with the situation. Participants were allowed to freely interact with the flight simulation, but were asked to only call up the DAS (with \textit{Recommendations} and \textit{Baseline}) or enter emergency mode (with \textit{Continuous Support} and \textit{Recommendations + Continuous Support}) when the incident happened. With \textit{Continuous Support} and \textit{Recommendations + Continuous Support}, pilots were allowed to freely interact with the normal flight mode before the incident.

Participants were asked to think aloud and clearly announce which airport they would divert to; they did not need to execute the diversion. The pilots were encouraged to think beyond what they saw in the DAS and consider whatever they would in a real flight, including which other stakeholders they would contact, like their company or air traffic control. They were further told that whatever is not shown in the table is not considered by the AI.

After the scenarios, we conducted semi-structured exit interviews with each participant to discuss their impressions of the system and how they used it. The interview guide is given in \autoref{sec:interview_guide}. Each study session took around 90 minutes, for which participants were paid 150~EUR ($\approx$ 164~USD) each. Again, this is a typical rate for pilots due to the difficulty of recruiting them. We recorded audio as well as the screen of the tablet running the DAS. The study was approved by \anon[the IRB at an anyonymized institution]{the Ethics Committee of the Faculty of Mathematics, Computer Science and Statistics at LMU Munich}.

We pilot-tested the study procedure with two of the focus group participants from \autoref{sec:focus_group} (see \autoref{sec:participant_details}, \autoref{tab:focus_participants}), whom we also paid 150~EUR each. Following the pilot test, we made small adjustments to the DAS user interface and to the third scenario. Most importantly, both pilot testers asked for preferences of the company in Scenario 3. We therefore decided that if participants would ask for this information, we would tell them that the airline would prefer a diversion to either Hanover, Berlin, or Düsseldorf. Hanover was the top recommendation of the AI, Düsseldorf was the third recommendation, and Berlin was not recommended due to slightly worse weather.

\subsection{Data Analysis}
We took participants' decisions and the time they needed for their decisions as outcome measures. As decision time, we took the time between calling up the system or entering emergency mode and announcing the decision. We further noted the reasons for the decisions from the think-aloud protocols to cover both outcome and process measures, as recommended by Roth~et~al.~\cite{roth_methods_2022}. 

To analyze the qualitative data, we transcribed the think-aloud protocols and exit interviews and coded both parts through thematic analysis~\cite{blandford_analysing_2016}. An initial round of open coding was conducted independently by two authors on four transcripts, one per study condition. The two authors then discussed their initial set of 269 low-level codes and consolidated them into 18 code groups. The first author coded the rest of the transcripts, extending and revising the initial coding scheme when necessary. All additional and revised low-level codes in this step continued fitting into the 18 code groups. Finally, we identified three themes that reflect the code groups. 

Additionally, we extracted usage patterns from the think-aloud protocols by reviewing the codes in their temporal order in each of the 96 decision instances (32 participants $\times$ 3 scenarios), referring back to the screen recordings where necessary for more context.

\section{Results}
\label{sec:results}
For our main study, we recruited 32 professional pilots through snowball sampling (two captains, rest first officers; two female, rest male; median age: 31 years (IQR 30--35); median flight hours: 2500 hours (IQR 2000--4000); median of 1.5 self-performed diversions (IQR 0--3); details in \autoref{sec:participant_details}, \autoref{tab:study_participants}). The pilots work in four German airlines, with past experience in eight additional airlines. 

We structure our results according to our research questions from \autoref{sec:research_questions}. Hereafter, we use the abbreviations \textit{Rec}, \textit{Cont}, and \textit{Rec+Cont} according to \autoref{fig:conditions} for the respective system variants. Participants are denoted with R$x$, C$x$, RC$x$, and B$x$ according to the system variant they used.

\subsection{RQ1: Workflow Integration}
\label{sec:results_usage}
We first describe the usage patterns we identified, followed by a comparison of how often they occurred across scenarios and system variants.

\subsubsection{Usage patterns}
\label{sec:results_usage_patterns}
All participants intuitively integrated the DAS into the FOR-DEC framework they are familiar with. We therefore distinguish the identified usage patterns on two overarching levels: the options considered by participants, and the strategies they used to decide between the options. The first directly mirrors the \textit{options} step in FOR-DEC, while the latter is a combination of the \textit{risks \& benefits} and the \textit{decision} steps, which were not always clearly distinguishable in the think-aloud protocols. We identified three distinct patterns for the options considered:
\begin{itemize}
    \item \textbf{O1---First few options}: Participants considered the first few options in the table. For \textit{Rec} and \textit{Rec+Cont}, these were the AI recommendations. For \textit{Cont} and \textit{Baseline}, these were the closest airports, since the table was sorted by time per default.
    \item \textbf{O2---First option}: Like the above, but participants only considered the very first instead of the first few options in the table.
    \item \textbf{O3---Self-generated options}: Participants generated their options independently from the order in the table, e.g. by sorting the table according to a column of interest, by looking for familiar airports, or by asking for company preferences. 
\end{itemize}
We further identified twelve strategies that participants used to decide between the options they considered. The first five are general strategies that could be used across all variants:
\begin{itemize}
    \item \textbf{S1---Narrow down}: Participants carefully compared the considered options, ruling out options one by one until one remains.
    \item \textbf{S2---Recognize best}: Participants immediately recognized the best among the considered options and focused on it without detailed comparison against the other options.
    \item \textbf{S3---Check if first option works}: In case participants considered only the first option, they checked whether there was anything against it.
    \item \textbf{S4---Confirm first option is best}: In case participants considered only the first option, they did a quick cross check to confirm that it was indeed the best, e.g. by glancing over the \textit{time to destination} to see if the first option was significantly closer than the next ones.
    \item \textbf{S5---Check one after another}: Participants checked whether there was anything against the first option. If yes, they checked the second one, and so on.
\end{itemize}
The seven remaining strategies capture how participants used the AI support elements available to them. The normal flight mode of the \textit{Cont} and \textit{Rec+Cont} variants was used in two different ways:
\begin{itemize}
    \item \textbf{S6---Use prepared plan}: Participants prepared a plan for a hypothetical emergency during normal flight. Even though they did not know what would happen, they could prepare on the level of e.g., \textit{``If something very urgent would happen, $x$ would be a good option.''} When the emergency happened, participants used this plan to quickly reach a decision.
    \item \textbf{S7---Refine situation awareness}: During normal flight, participants were aware of the general situation, such as the weather around them. When the emergency happened, participants did not review this information again, but only supplemented it with situation-specific information like the distance to the next hospital in Scenario 2.
\end{itemize}
The color highlights, which served as transparency for recommendations and as local hints for continuous support, were also used in two ways:
\begin{itemize}
    \item \textbf{S8---Look for options without highlights}: Participants reviewed highlights for their relevance, but gravitated toward options without highlights. This strategy was used for choosing among the considered options as well as for deciding which options to consider in the first place. For instance, participants looked at \textit{time to destination} and quickly excluded those options with red highlights, which indicated that these airports were too far away.
    \item \textbf{S9---No highlights as confirmation}: Participants had a decision in mind and felt confirmed if that option had no highlights. The difference to above is that there, the highlights were used as cue to guide the decision-making, while here, participants considered the highlights as a second opinion to their own independent reasoning.
\end{itemize}
Lastly, participants employed three distinct strategies for using recommendations:
\begin{itemize}
    \item \textbf{S10---Recommendation as confirmation}: Participants had a decision in mind and felt confirmed if that option was also recommended by the DAS. Some pilots using this strategy explicitly ignored the recommendations initially to reach a decision on their own first.
    \item \textbf{S11---Recommendation as fallback}: Participants first checked their self-generated options. When they were not satisfied with any of them, they chose the AI-recommended option.
    \item \textbf{S12---Negotiate}: Participants edited the criteria to see how it affected the recommendations. This strategy was triggered by one of two reasons, or a combination of both: Some pilots disagreed with the pre-defined criteria, while others were surprised that their favored option was not recommended. In the latter case, they tried to align the recommendations with their own opinion by removing or relaxing criteria they deemed uncritical. Those that successfully aligned the recommendations to their favored option took this as confirmation. One pilot did not succeed, which triggered further considerations in his decision-making.
\end{itemize}

\subsubsection{Differences between scenarios and support paradigms}
\label{sec:results_usage_differences}
\begin{figure*}[thb]
    \includegraphics[width=\textwidth]{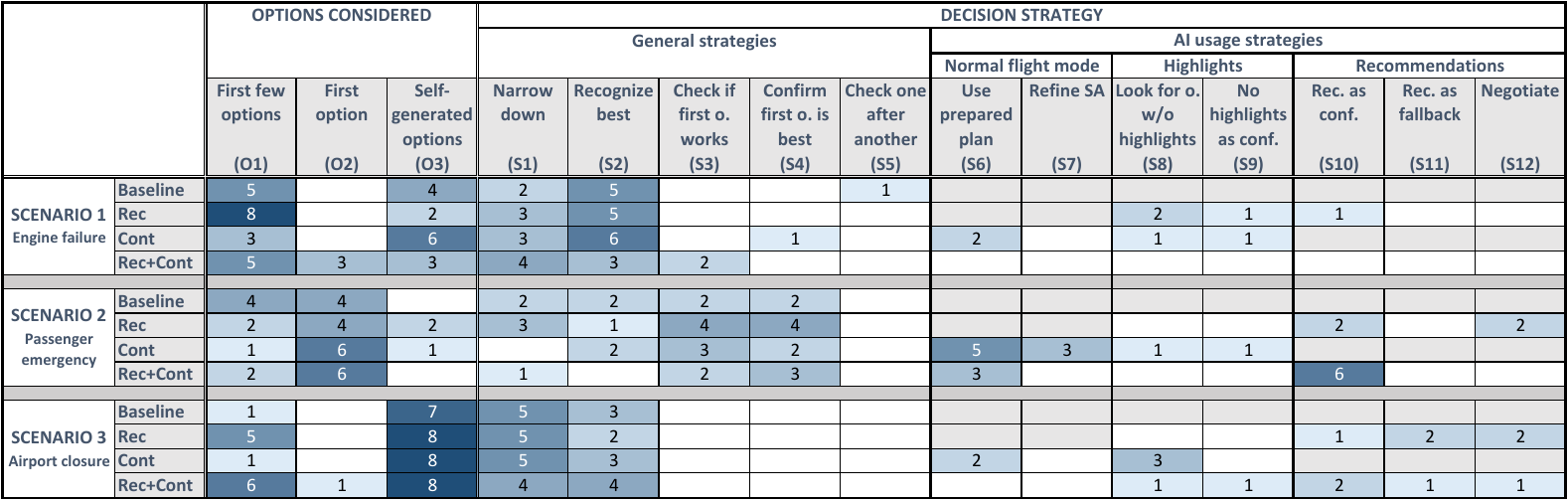}
    \caption{Usage patterns across conditions and scenarios. The counts exceed the number of participants because individual usage patterns are not mutually exclusive and participants typically employed more than one pattern. Grayed-out fields mean that the patterns are not applicable to the respective system variant.}
    \label{fig:usage_patterns}
    \Description{A very complex table of the different usage patterns, with shades of blue to visualize the frequency of pattern occurances. Columns hold the patterns, categorized by options considered and decision strategy. The latter is further split into general strategies and AI usage strategies. The latter is again further split into normal flight mode, highlights, and recommendations. The rows hold the four conditions, each three time for the three scenarios.}
\end{figure*}
\autoref{fig:usage_patterns} gives an overview of the occurrences of each usage pattern across conditions and scenarios. Note that the real occurrences could be higher, since participants might have used certain strategies without any revealing verbalizations. Especially some of the AI usage strategies were more subconscious and therefore less likely to be verbalized. 

The most obvious difference is between the options considered in different scenarios, where participants tended to consider the \textit{first few options} (O1) in Scenario 1 and only the \textit{first option} (O2) in Scenario 2, while they based their decision more heavily on \textit{self-generated options} (O3) in Scenario 3. The difference was mostly due to the varying time criticality, as several participants explained during the exit interviews. The engine failure in Scenario 1 was of medium time criticality, the passenger emergency in Scenario 2 was extremely time-critical, while the airport closure in Scenario 3 was not time-critical at all.

Besides these scenario differences, there are also differences between the DAS variants. These are most salient in Scenario 2, where participants in the \textit{Cont} and \textit{Rec+Cont} groups benefitted the most from the possibility of detailed pre-planning with the normal flight mode. As a result, pilots in these groups could \textit{use their prepared plan} (S6) or use \textit{recommendations as confirmation} (S10) for their plan, indicating forward reasoning by these participants. Note that pilots using the \textit{Baseline} and \textit{Rec} variants also tried to establish SA during normal flight, as they would in reality, but could only rely on the much more limited possibilities offered by X-Plane. Consequently, when the emergency happened, pilots in the \textit{Rec} group were more likely to take the system recommendation as starting point for review (S3, S4), which indicates backward reasoning. In the absence of AI support that helps to identify the best option, participants using the \textit{Baseline} variant had a stronger tendency to review the \textit{first few options} (O1) rather than only the \textit{first option} (O2).

In Scenarios 1 and 3, the benefits of pre-planning during normal flight in \textit{Cont} and \textit{Rec+Cont} are less apparent. This can be explained by a disruption between normal flight and emergency decision-making in both scenarios. In Scenario 1, the disruption happened because pilots first handled the engine failure before entering the diversion decision, a procedure that took several minutes during which the position of the aircraft changed significantly. After securing the engine, pilots therefore had to slightly re-orient and could not directly use their SA from the normal flight. In Scenario 3, the disruption occurred since pilots mostly prepared for a time-critical emergency requiring a quick landing. What happened instead was a situation where it was more important to find a suitable airport near the destination so that participants' preparations were not applicable. 

Nevertheless, a difference is still observable between \textit{Rec} and \textit{Cont} in Scenario 1, indicating that continuous support encourages forward reasoning, while recommendations prompt backward reasoning. All participants using the \textit{Rec} variant took the system recommendations as starting point (O1), while participants in the \textit{Cont} group tended to consider \textit{self-generated options} (O3), suggesting that they could build on their normal flight SA to some extent, despite the disruption of handling the engine failure. Furthermore, some participants in the \textit{Rec+Cont} group limited their options to the system's top recommendation (O2), which could have been because it conformed to their impressions of the available options from the normal flight phase.

In Scenario 3, a stark contrast is apparent between \textit{Baseline} and \textit{Cont} on the one hand---which did not have AI recommendations---and \textit{Rec} and \textit{Rec+Cont} on the other hand. Almost all participants in this scenario considered \textit{self-generated options} (O3), mostly by asking for company preferences and by looking for familiar airports. In addition, those participants who had recommendations available to them also mostly considered them (O1), indicating a strong influence of the recommendations. However, from the think-aloud protocols, we were not able to tell whether they influenced participants' reasoning differently between \textit{Rec} and \textit{Rec+Cont}, given that the observable recommendations usage patterns are very similar (S10, S11, S12). Still, the observed usage patterns presented in \autoref{fig:usage_patterns} suggest that our hypothesis holds true according to which recommendations prompt backward reasoning, while continuous support facilitates forward reasoning, even though the latter can be derailed by disruptions between normal flight and emergency decision-making.

\subsection{RQ2: Overreliance}
\label{sec:results_outcomes}
Decision outcomes were quite uniform for the first two scenarios. For Scenario 1, this was to be expected, since the scenario was designed to be rather unambiguous for the reason given in \autoref{sec:scenario_1}. All participants except for C5 (Lyon) and B8 (Paris Orly) decided for Paris Charles de Gaulle in Scenario 1, which was also the system's top recommendation.
However, in Scenario 2, all participants except for B5 (Faro) chose Casablanca, which ran contrary to our expectation given the discussion of our focus group participants, as described in \autoref{sec:scenario_2}. Some participants in the main study, especially those who had flown in the region themselves before, emphasized that the bias against diverting to Africa is common among their colleagues, but unwarranted in this case. 14 participants across all conditions did mention that they would generally prefer to divert to Faro for the reasons also discussed by the focus group participants. In the end, the fact that the emergency was urgent and that Casablanca was around ten minutes closer tipped the scales for all of these pilots. The AI elements---which favored Casablanca---did not seem to have an effect on this decision, given that almost all participants in the \textit{Baseline} group also decided for Casablanca.

As intended, Scenario 3 was less clear for participants, as shown in \autoref{tab:scenario_3_decisions}. Of most interest in terms of overreliance was how many participants decided for Hanover without considering the traffic situation, as explained in \autoref{sec:scenario_3}. While some participants did rule out Hanover themselves because of the traffic, other pilots either said they would ask air traffic control about the traffic density, or they considered traffic but expected it to be not too dense to land there. We told these participants that Hanover was already running out of capacity, as air traffic control would do in reality. We did so since we were not interested in how pilots would assess the traffic situation, but only whether they would consider it at all. The DAS does not include this information, so we were interested in whether this would lure pilots into overlooking this factor, or whether they would think beyond the limits of the system.
\begingroup
\renewcommand{\arraystretch}{1.2}
\renewcommand\cellset{\renewcommand\arraystretch{0.8}\setlength\extrarowheight{0pt}}
\begin{table}[tbh]
    \centering
    \caption{Diversion decisions across conditions in Scenario 3. Hanover was to be avoided due to heavy traffic, which was not considered by the system.}
    \label{tab:scenario_3_decisions}
    \begin{tabular}{l l | p{0.1\textwidth}<{\centering} p{0.1\textwidth}<{\centering} p{0.1\textwidth}<{\centering} p{0.1\textwidth}<{\centering}}
        \toprule
        & & \textbf{Baseline} & \textbf{Rec} & \textbf{Cont} & \textbf{Rec+Cont}\\
        \midrule
        \makecell[tl]{To be\\avoided} & \makecell[tl]{\textbf{Hanover}\\\textit{AI recommendation \&}\\\textit{company preference}} & 1/8 & 6/8 & 4/8 & 3/8\\
        \midrule
        Other & \makecell[tl]{\textbf{Berlin}\\\textit{company preference}} & 7/8 & - & 4/8 & 3/8\\
        & \makecell[tl]{\textbf{Düsseldorf}\\\textit{company preference}} & - & 1/8 & - & 1/8\\
        & \textbf{Cologne} & - & - & - & 1/8\\
        & \textbf{Frankfurt} & - & 1/8 & - & -\\
        \bottomrule
    \end{tabular}
\end{table}
\endgroup

\begin{figure*}[thb]
    \includegraphics[width=0.55\textwidth]{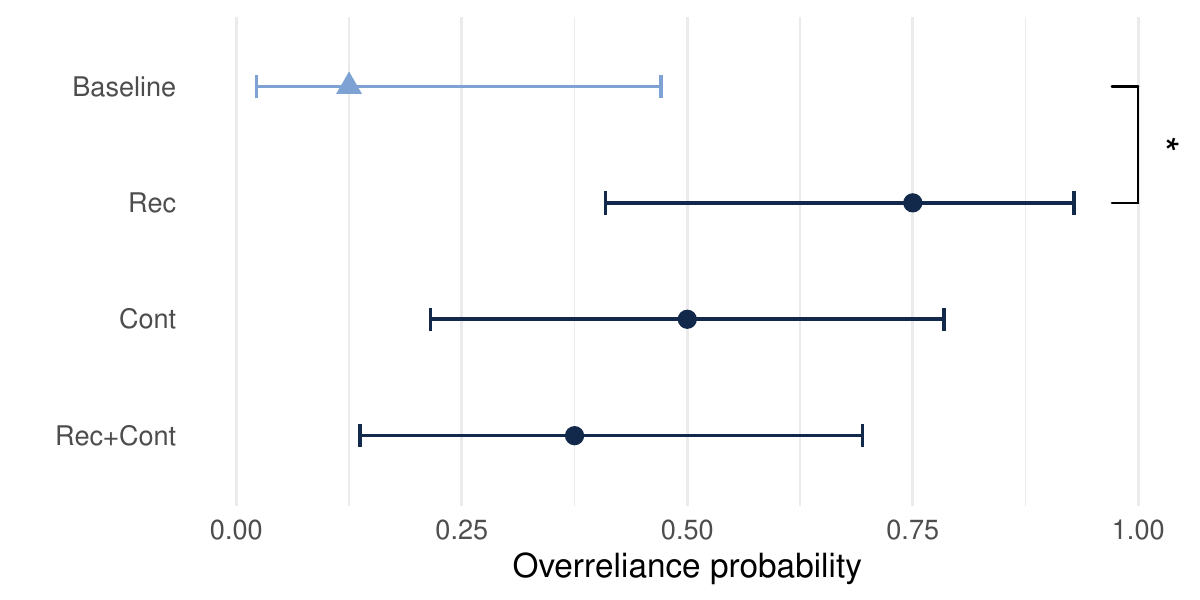}
    \caption{Probability of overreliance across conditions, where choosing Hanover in Scenario 3 without considering traffic conditions is counted as overreliance. Error bars denote 95\% confidence intervals, estimated using the Wilson score interval method. (*) denotes $p < 0.05$.}
    \label{fig:overreliance}
    \Description{Point plot of overreliance occurrences, with conditions on y-axis and overreliance probability on x-axis. The probabilities are: Baseline 0.125, Rec+Cont 0.375, Cont 0.5, Rec 0.75. The error bars are very long, spanning around 0.4 for each condition.}
\end{figure*}

\autoref{fig:overreliance} shows how the three AI system variants affected the probability for the overreliant behavior of choosing Hanover without considering traffic, as compared to the \textit{Baseline} variant. Since we assumed that adding AI would increase the probability for overreliance, we performed one-sided Fisher's exact tests to compare each AI variant with \textit{Baseline}. We further report relative risks (RR) compared to \textit{Baseline} with Wald normal approximation confidence intervals as effect sizes. Consistent with our hypothesis, there was a statistically significant increase of overreliance probability for \textit{Rec} ($RR=6$, 95\% CI [0.92, 39.18], $p=0.02<0.05$). For \textit{Cont} ($RR=4$, 95\% CI [0.56, 28.40], $p=0.14$) and \textit{Rec+Cont} ($RR=3$, 95\% CI [0.39, 23.07], $p=0.28$), the increase was smaller and not statistically significant, though these latter results have to be interpreted carefully given the small sample sizes and the resulting large confidence intervals. However, it appears like even without recommendations, pilots using the \textit{Cont} variant could still be biased by the hints: Some airports in Scenario 3, including Berlin, had a braking action of ``medium to good'', i.e., slightly wet runways, which the system highlighted in red. With the \textit{Baseline} variant, which did not have highlights, participants acknowledged the wet runways with a short comment that it was uncritical; some even did not verbalize any thoughts about it at all. By contrast, with the AI variants, many participants found the red highlights hard to ignore, even though they deemed the braking action uncritical: \textit{``To me, red is always intuitively, it doesn't look so good. That's why I looked a bit at Hanover, even though `medium to good' is fine.''}~(C8). This behavior is also reflected by the usage patterns in \autoref{fig:usage_patterns}, where some \textit{Cont} participants looked for options without highlights (S8).

In addition to the decisions, we also noted the reasons behind them. \autoref{tab:scenario_3_reasons} gives an overview of the reasons why non-overreliant participants decided against Hanover, showing that traffic density was indeed the reason for most of them. Some participants decided against Hanover but not due to traffic. These pilots---all of whom saw no recommendations---favored Berlin for other reasons and focused on it without discussing Hanover in depth. Interestingly, the pilots in the \textit{Rec} group who thought of traffic were two of the three pilots in that group who ignored the recommendations (i.e., not used the pattern O1 in \autoref{fig:usage_patterns}) to make their own independent thoughts. This suggests that forward reasoning was important for pilots to be able to think beyond the limits of the system.
\begingroup
\renewcommand{\arraystretch}{1.2}
\renewcommand\cellset{\renewcommand\arraystretch{0.8}\setlength\extrarowheight{0pt}}
\begin{table}[tbh]
    \centering
    \caption{Reasons for not diverting to Hanover in Scenario 3. Participants could give more than one reason.}
    \label{tab:scenario_3_reasons}
    \begin{tabular}{l | p{0.1\textwidth}<{\centering} p{0.1\textwidth}<{\centering} p{0.1\textwidth}<{\centering} p{0.1\textwidth}<{\centering}}
        \toprule
        & \textbf{Baseline} & \textbf{Rec} & \textbf{Cont} & \textbf{Rec+Cont}\\
        \midrule
        Traffic in Hanover & 5/7 & 2/2 & 3/4 & 5/5\\
        Berlin is closer to current location & 2/7 & - & 1/4 & -\\
        \makecell[tl]{Berlin has better connections for\\crew to return home} & - & - & 1/4 & -\\
        \bottomrule
    \end{tabular}
\end{table}
\endgroup

\subsection{RQ3: Decision Time}
\begin{figure*}[thb]
    \begin{subfigure}[b]{0.4\textwidth}
        \centering
        \includegraphics[width=\textwidth]{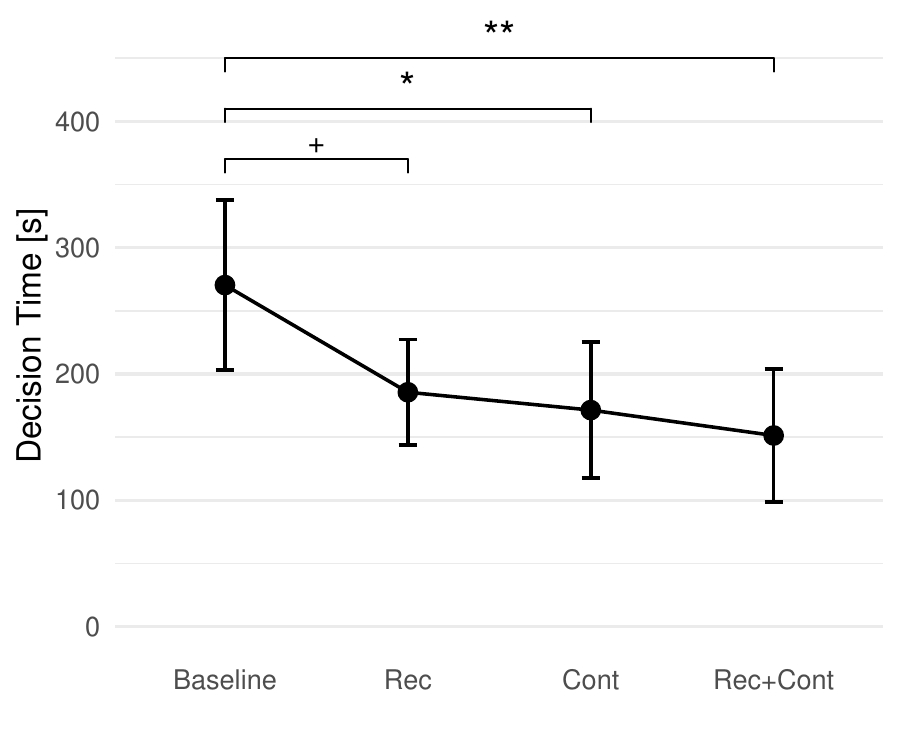}
        \caption{Scenario 1 (engine failure)}
        \label{fig:scenario_1_time}
        \Description{Line plot with error bars. Decision time in seconds on y-axis, from 0 to 400, and the four conditions on x-axis. Baseline has clearly the highest decision time (around 275 seconds), followed by Rec, Cont, and Rec+Cont, in descending order, which are all relatively close to each other (all between 150 and 190 seconds).}
    \end{subfigure}
    \begin{subfigure}[b]{0.4\textwidth}
        \centering
        \includegraphics[width=\textwidth]{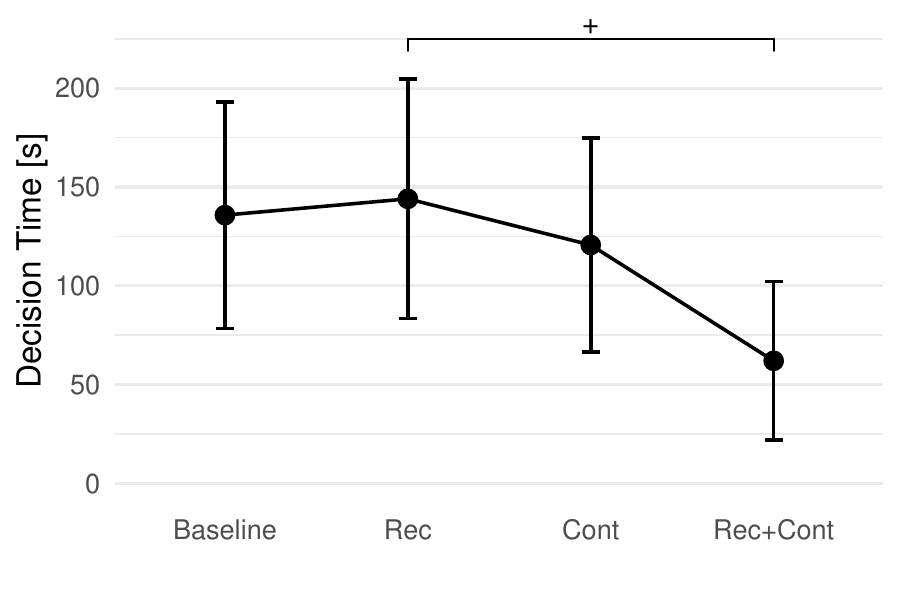}
        \caption{Scenario 2 (passenger emergency)}
        \label{fig:scenario_2_time}
        \Description{Line plot with error bars. Decision time in seconds on y-axis, from 0 to 200, and the four conditions on x-axis. Rec+Cont has clearly the lowest decision time (around 60 seconds), followed by Cont, Baseline, and Rec in ascending order, which are all relatively close to each other (all between 125 and 150 seconds).}
    \end{subfigure}
    \begin{subfigure}[b]{0.4\textwidth}
        \centering
        \includegraphics[width=\textwidth]{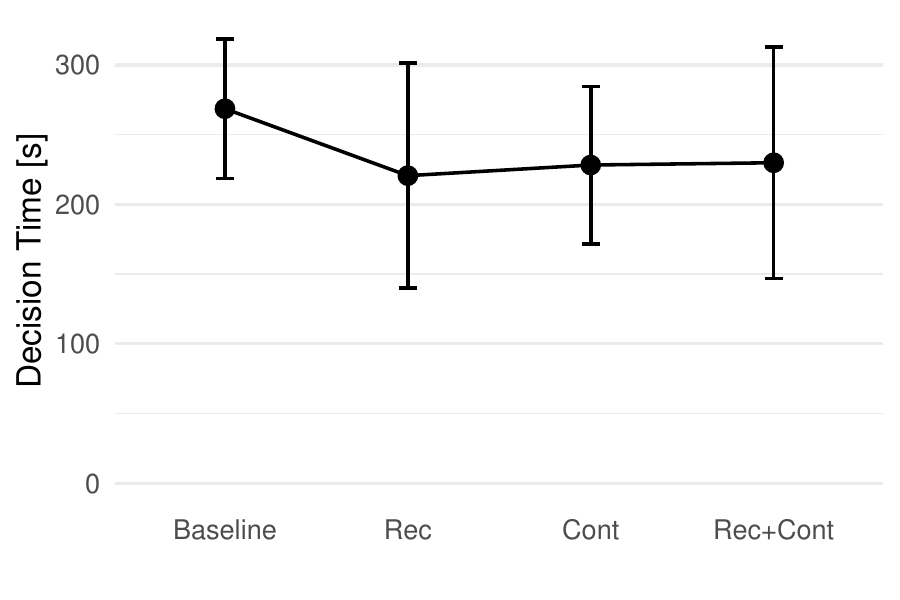}
        \caption{Scenario 3 (airport closure)}
        \label{fig:scenario_3_time}
        \Description{Line plot with error bars. Decision time in seconds on y-axis, from 0 to 300, and the four conditions on x-axis. Baseline has by a small margin the highest decision time (around 270 seconds). Rec, Cont, Rec+Cont, in ascending order, are close to each other (all around 225 seconds).}
    \end{subfigure}
    \caption{Decision time across conditions for all three scenarios. Error bars represent 95\% confidence intervals. (+) denotes $p < 0.1$, (*) denotes $p < 0.05$, and (**) denotes $p < 0.01$.}
    \label{fig:decision_times}
\end{figure*}

The differences in usage patterns discussed in \autoref{sec:results_usage_differences} also resulted in different patterns in decision times, as shown in \autoref{fig:decision_times}. We therefore analyzed each scenario separately through an analysis of variance (ANOVA). We further used Tukey's HSD test for post-hoc tests and report both mean differences in seconds and Cohen's $d$ as effect sizes. 
\label{sec:results_time}

In Scenario 1, we found statistically significant differences in decision times at the $p < 0.01$ level ($F(3,28) = 5.15, p = 0.006$), with a large effect size of $\omega^2 = 0.28$. Post-hoc tests revealed that all three AI system variants led to significantly faster decisions than \textit{Baseline}: \textit{Rec} was significantly faster than \textit{Baseline} at the $p < 0.1$ level ($\Delta = 84.88$ s, 95\% CI [-4.30 s, 174.05 s], $d = 1.30$, 95\% CI [-0.20, 2.80], $p = 0.066$), \textit{Cont} at the $p < 0.05$ level ($\Delta = 99.00$ s, 95\% CI [9.83 s, 188.17 s], $d = 1.52$, 95\% CI [-0.02, 2.05], $p = 0.025$), and \textit{Rec+Cont} at the $p < 0.01$ level ($\Delta = 119.13$ s, 95\% CI [29.95 s, 208.30 s], $d = 1.82$, 95\% CI [0.25, 3.40], $p = 0.006$). The effect sizes can be considered large in all three comparisons. The differences between the AI variants were not statistically significant.

In Scenario 2, decision times differed significantly at the $p < 0.1$ level ($F(3,28) = 2.66, p = 0.067$) with a medium effect size of $\omega^2 = 0.135$. Post-hoc test results showed that \textit{Rec+Cont} was significantly faster than \textit{Rec} at the $p < 0.1$ level with a large effect size ($\Delta = 82.00$ s, 95\% CI [-5.61 s, 169.61 s], $d = 1.28$, 95\% CI [-0.22, 2.78], $p = 0.073$). Other post-hoc comparisons were not statistically significant. The most notable difference to Scenario 1 was that the \textit{Baseline} variant was similarly fast as \textit{Rec} and \textit{Cont}. This effect can be explained by the difference in time criticality, as discussed in \autoref{sec:results_usage_differences}. In Scenario 1, pilots compared several options in greater detail, which was facilitated by AI support; whereas in Scenario 2, they tended to only check the nearest airport to speed up the decision. This short check could be performed quickly even without AI support.

There were no statistically significant differences in decision times in Scenario 3 ($F(3,28) = 0.544, p = 0.66)$, with an effect size close to zero. This again can be explained by the factor of time criticality. Participants had no time pressure in Scenario 3 and hence took their time for in-depth comparisons, irrespective of which system variant they used.

Overall, the most notable observation was that \textit{Rec} was surprisingly not faster than \textit{Cont}; it even tended to be slightly slower in the first two scenarios, though not significantly. Interestingly, the combination \textit{Rec+Cont} did lead to a statistically significant speedup in Scenario 2. This is consistent with the usage patterns in \autoref{fig:usage_patterns}: Many participants in this group already prepared during normal flight for Casablanca in case of a time-critical emergency. When the emergency happened, the system confirmed their prepared plan (S10, forward reasoning), reducing the time to verify the plan. With \textit{Rec}, participants got the same recommendation, but since they were less prepared, they had to verify the recommendation first (S3, S4, backward reasoning), which consumed a significant amount of time. The same trend can be observed in Scenario 1, though less pronounced, likely due to less time pressure and the disruption of handling the engine failure, as discussed in \autoref{sec:results_usage_differences}.

\subsection{RQ4: Pilots' Perspectives}
\label{sec:results_perspectives}
We identified three major themes in our thematic analysis of the exit interviews and occasional comments during the think-aloud sessions.

\subsubsection{Greatest added value: making a lot of information quickly accessible}
For participants, the greatest added value of the DAS was not the AI features per se, but rather the quick overview of a large amount of relevant information, as explained by 30 out of 32 participants. Today, pilots must gather information from various sources, which is time-consuming and error-prone. Our DAS improves on the current state even in the \textit{Baseline} version by presenting all information at a glance in the table. Several participants mentioned that the table is basically what pilots in some airlines fill out by hand as part of FOR-DEC. Having it directly available speeds up decision-making and enables better decisions by reducing workload and providing more complete information:
\begin{quote}
    \textit{``If you have time, you can work it all out in half an hour. But if you have a time-critical failure, [...] you need a lot of experience because you don't have the time to work it all out. You can spend hours doing FOR-DEC, but you only have fuel for half an hour. And then you have to make an intuitive decision based on your gut feeling. These are often not the best decisions.''}~(C2)
\end{quote}

But AI features do contribute to the quick overview. While having all information at a glance is helpful, it is also overwhelming, as criticized by six participants. Pilots see the importance of AI in reducing clutter by surfacing relevant and hiding or de-emphasizing less relevant information. 14 pilots stressed that recommendations and ranking by AI evaluation help them to quickly focus on good options: \textit{``I think this pre-filtering is important. And afterwards, the human can still fine-tune it a little.''}~(RC5). Also 14 participants explained how the color highlights help them to quickly identify potential limitations at airports: \textit{``It's quite clever if you can see right away, do I even need to look at it or not?''}~(RC2). Participants also suggested how AI could further reduce clutter, e.g. by hiding very bad options, or by evaluating weather as a whole and only displaying noteworthy weather aspects. In our system, all weather components are shown in the table and evaluated individually, creating a lot of redundancy: \textit{``A little less information, visibility 10,000 or more, that's redundant information, that doesn't help me. So more with the question, what information creates added value?''}~(R5).

Two participants cautioned against careless implementations of AI. They pointed out that especially the value of recommendations may be undermined by the need to double-check them, which could slow pilots down: \textit{``You are inclined to question the computer, and you also want to decide for yourself, but at the same time you want to use everything, and then you'll probably need ages to figure it all out.''}~(B8). This view is consistent with the decision times in \autoref{sec:results_time}, where the \textit{Rec} variant turned out slower than expected.

\subsubsection{Tension between more system intelligence and more bias}
\label{sec:tension}
While pilots welcomed AI support to surface relevant information, they were also concerned that too obtrusive AI could bias them toward subpar decisions: 
\begin{quote}
    \textit{``The more is presented to you and processed for you, the less you are actively involved. On the one hand, it helps a lot. On the other hand, you have to be careful whether all of this is exactly what you actually want.''}~(C5)
\end{quote}
\begin{quote}
    \textit{``It's a fine line as to whether you do too much.''}~(B1).
\end{quote}
This tension was especially apparent with recommendations: 17 participants found recommendations helpful or suggested to add them if their system variant did not have them, while eleven participants rejected them or found them not helpful. Those who rejected recommendations felt that they remove pilots from the decision-making:
\begin{quote}
    \textit{``I find it a bit difficult that the decision is given to you immediately like this [snaps fingers], and then you immediately go, ah okay, if it says so, then we'll take Hanover.''}~(RC4).
\end{quote}
\begin{quote}
    \textit{``What I would not like is for the AI to simply say `Fly to Berlin!' or something, because then you don't know exactly where it's coming from. The human is the master in the system, the AI has to be subordinate and perform supporting tasks.''}~(B5)
\end{quote}
Other participants saw no threat in recommendations to their agency, emphasizing that it is their job to monitor and question the system.

Hints and highlights were seen as less problematic and accepted by all participants, as they were perceived as \textit{``less patronizing''}~(R5): \textit{``I have the feeling that I'm using the system, not that the system is using me.''}~(C2). However, seven pilots stressed that the highlights must make sense to be useful. Pilots especially disagreed with the red highlights of the ``medium to good'' braking action in Scenario 3, as described in \autoref{sec:results_outcomes}.

Participants discussed several ways how AI may be introduced without removing pilots from the decision-making. For one, nine pilots noted that continuous support helps them to familiarize with the system and improve their SA, which also makes using the system in an emergency easier:
\begin{quote}
    \textit{``It's important to be familiar with the system, because if you only use it in an emergency, it's like, oh yeah, is that right? If you're always plotting where you are, you already know a bit of what it says and you don't see, ah okay, it's red, it doesn't work.''}~(C8)
\end{quote}
Moreover, participants had several ideas for alternative forms of AI support. Some suggested that the system only marks whether airports are suitable for a diversion or not, instead of making specific recommendations. Another idea was to display recommendations only after pilots have independently reached a decision, which some pilots likened to their collaborative decision-making:
\begin{quote}
    \textit{``It's like with our CRM\footnote{\textit{Crew Resource Management}, procedures for effective communication and decision-making to prevent human error.}, when we work together, I have to be careful not to say as captain `I think that airport is great, that's where we're flying to now, or would you mind?' Then of course he [the first officer] says `No, let's do it.' That's why I always have to keep it open and not say what I've been thinking. Let the other one say it so that there is redundancy, or perhaps have my own mistakes pointed out.''}~(C5)
\end{quote}
Another suggestion by several participants was that pilots have to manually define their criteria before they get recommendations. The system could also step in with suggestions for additional criteria that the pilot may have forgotten. RC2 proposed that AI could be used to evaluate the airports according to high-level categories like weather or operations. Pilots could then sort the airports by these high-level categories, rather than by low-level criteria like in our system. C8 advocated that the system should be limited to hard facts and leave soft factors to pilots:
\begin{quote}
    \textit{``The wind won't change, that's for sure. The fuel won't change either. But I still have to see where the company wants us to go. Where do we perhaps know our way around? [...] So these soft factors, if they were included, I think that would be a bit like taking the decision away.''}~(C8)
\end{quote}
However, opinions differed on this last suggestion, as B2 argued for the opposite and said that the system should be able to judge when a situation is so critical that it is acceptable to land somewhere \textit{``even if I am now busting a limit.''}~(B2). The system should be able to \textit{``classify, how serious is the incident for the danger of all people on board or for the aircraft?''}~(B2).

While the discussions were predominantly about the risk of system-induced biases, four pilots brought the complementary perspective into the conversation, pointing out that AI could also help mitigate human errors. AI could for instance highlight good but less familiar options, and does not overlook things under time pressure.

\subsubsection{Transparency and control: need for appropriation}
Lastly, pilots discussed several ways in which they require transparency and control. Some participants wondered about how the AI works, e.g., why a certain recommendation is given (three participants), how the criteria are weighted (four participants), or why certain information is highlighted (four participants). However, the much more prevalent issue was what exactly the system considers for its evaluations (17 participants): Does \textit{time to destination} include the time required for the approach? Does \textit{fuel at destination} include final reserves? Does \textit{stop margin} consider performance limitations after a technical failure? The predominant question was therefore not ``How does the AI work?'', as assumed by most research in explainable AI, but rather ''How does the information fit my intention?''.

Pilots further expressed the requirement to be able to control the AI. Five participants explicitly valued the option to edit the recommendation criteria to engage with the AI in negotiation patterns, as described in \autoref{sec:results_usage_patterns}. Moreover, participants made further design suggestions to enable more control, such as options to pin airports to the top of the table and to manually hide airports, or to search for airports that are not in the table. R3 further suggested to add an extra column that pilots can fill themselves, for edge cases that are not covered by the system criteria.

The common theme behind both transparency and control was that pilots want to be able to appropriate the system to fit their intention and momentary information needs. Most of the transparency and control requests were comments and questions interposed during the think-aloud sessions, where participants had an intention while using the system and asked for clarification to understand if the system would fit their intention.

\section{Discussion}
We discuss the takeaways of our results in \autoref{sec:beyond_rec} and \autoref{sec:beneficial_rec} and bring them together in \autoref{sec:process_oriented_support}. We close with limitations and future work in \autoref{sec:limitations}.

\subsection{There Is More to Decision Support than Giving Recommendations and Explanations}
\label{sec:beyond_rec}
The common, but often tacit design goal for AI decision support is to solve the task for users, creating a redundant rather than complementary role to the human. While it is popular to speak of ``human-AI collaboration'' in AI-assisted decision-making research~\cite{schmidt_calibrating_2020,hemmer_human-ai_2023}, users may perceive this recommendation-centric support more as ``human-AI competition''~\cite{zhang_rethinking_2024}. Some of our participants expressed similar concerns regarding recommendations.

Like previous studies involving real-world tasks with experts, our participants' views suggest a shift in the design goal from solving the entire decision task for users to addressing their primary pain points. For the tumor assessment use case studied by Lindvall~et~al.~\cite{lindvall_rapid_2021}, pathologists' biggest challenge was to find small tumorous regions in huge images. In the case of sepsis diagnosis studied by Zhang~et~al.~\cite{zhang_rethinking_2024}, physicians wanted to know which lab tests to order, when. In our case of diversions, pilots would benefit the most from a quick overview of a large amount of relevant information. This shift in design goal enables much more diverse uses for AI than merely giving end-to-end recommendations, as exemplified by the various suggestions by our participants. 

The challenge shifts from giving best possible recommendations and explaining them, to serving pilots' momentary information needs. Transparency is consequently not only---or maybe not even primarily---required to calibrate users' reliance on the recommendation, as is the current focus in AI-assisted decision-making research~\cite{fok_search_2024}. Rather, the role of transparency is to make visible how well the information fits pilots' current intention. 

Fine-grained control is further necessary to allow pilots to effectively cater the information to their intentions when the system-inferred information does not perfectly fit. This prevents the all-or-nothing situation often created by typical recommendation-centric support: Due to the closed nature of end-to-end recommendations, decision makers usually only have the choice to fully accept or reject the recommendation\footnote{At least in classification tasks. In regression tasks, users can give the AI recommendation a more continuous weight in their decision-making~\cite{poursabzi-sangdeh_manipulating_2021,green_principles_2019}.}, which is also how users' reliance behavior is often modeled~\cite{schemmer_appropriate_2023}. However, as observed by Sivaraman~et~al. with clinicians~\cite{sivaraman_ignore_2023}, decision makers' reliance behavior can be much more nuanced when they are given the opportunity. In their case, the AI recommendation consisted of several aspects, allowing clinicians to adopt some aspect while overruling another. With our DAS, pilots could go one step further by editing the recommendation criteria. This led to some productive uses of the system in instances where the pilot would have simply ignored or rejected the system had the controls not been available.

\subsection{Recommendations Have to Be Embedded into Forward Support to Be Beneficial}
\label{sec:beneficial_rec}
According to our results, recommendation-centric support significantly constrains pilots from thinking beyond the limits of the system, which in line with previous work~\cite{jacobs_how_2021,smith_brittleness_1997,bussone_role_2015} leads to more overreliance. Recommendations on their own also surprisingly did not lead to faster decisions than continuous support, as the need to review the recommendations canceled out the gains from surfacing good options. Lastly, the use of recommendations was highly disputed among participants. 

But while our results suggest that recommendations may not be the most valuable use of AI for diversion assistance, we did find benefits of providing recommendations. First, recommendations can serve as confirmation, which can accelerate decisions in time-critical situations, which was particularly evident in combination with continuous support in Scenario 2. Recommendations can further benefit decision-making when they challenge pilots' own ideas and trigger further considerations. This was more apparent in the interview statements, but could also be observed in one instance during the think-aloud sessions, namely with the participant that tried without success to align the recommendations with his own idea, as described in \autoref{sec:results_usage_patterns}. 

Note that both beneficial uses of recommendations require forward reasoning, as pilots must independently come up with a favored option to be confirmed or challenged. We found continuous support to be effective for promoting forward reasoning, but it lacked robustness against disruptions between normal flight and abnormal situations. Following such disruptions, pilots tended to revert to backward reasoning. To ensure reliable forward reasoning when presenting recommendations, additional measures are necessary, with potential suggestions provided by our participants.

\subsection{From Recommendation-Centric to Process-Oriented Support}
\label{sec:process_oriented_support}
\begin{figure*}[thb]
    \includegraphics[width=0.95\textwidth]{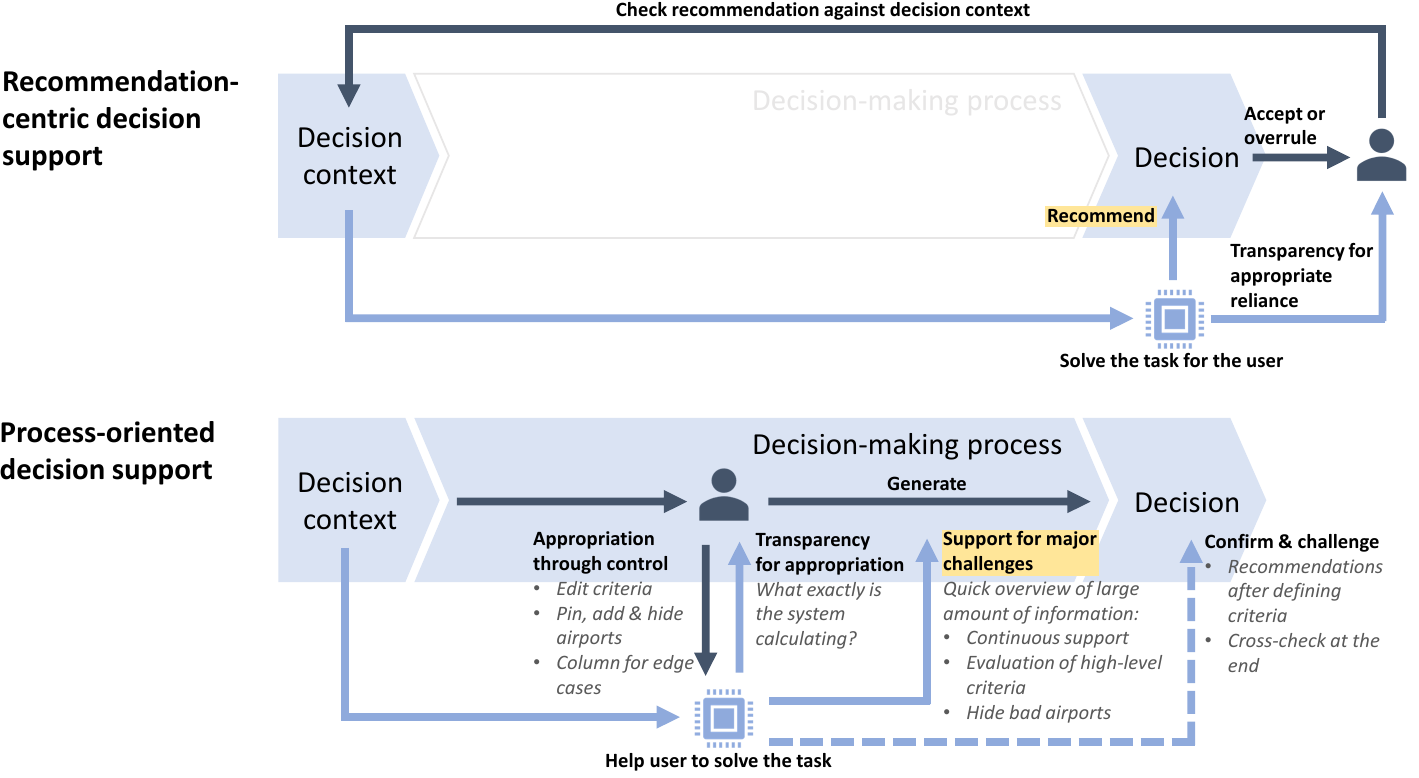}
    \caption{Comparison of recommendation-centric and process-oriented decision support. The yellow highlights signal the core functionality of the AI in both paradigms. The dashed arrow indicates that decision recommendations are optional in process-oriented support. The italic text lays out how process-oriented support can be implemented concretely for the diversion use case, based on our results and participants' suggestions.}
    \label{fig:framework}
    \Description{Schematic overview of recommendation-centric decision support on top, and process-oriented decision support on the bottom. In the former, the user is outside of the decision-making process, while in the latter, the user is within the decision-making process.}
\end{figure*}

Bringing together the discussed aspects, we propose \textit{process-oriented decision support} as a promising alternative framework to design AI decision support that is not centered around end-to-end recommendations. \autoref{fig:framework} shows how process-oriented support compares to recommendation-centric support, and how it applies to the diversion use case through both continuous support and participants' design suggestions. We consider the framework to be applicable not only for diversion assistance but more generally for AI support of high-stakes decisions. We argue that process-oriented support is particularly useful for complex decisions where human expertise and context knowledge is crucial, but hard to combine with end-to-end recommendations, such as in healthcare~\cite{zhang_rethinking_2024}, social work~\cite{kawakami_why_2022}, or sales~\cite{blomberg_acting_2018}. The key difference to recommendation-centric support is the shift from trying to solve the task \textit{for} users through end-to-end recommendations, to \textit{helping} users to solve the task by addressing the challenges in their decision-making process. These challenges have to be determined for each application through user research. In our diversion use case, pilots' main challenge was to gather the information they need to make a decision.

This shift in design goal has several implications. First, recommendation-centric support removes users from the decision-making process. It pushes users into backward reasoning, with the recommended end result as the starting point, leading to the lack of cognitive engagement and hence inappropriate reliance observed in prior work~\cite{bucinca_trust_2021,gajos_people_2022}. This is not only an issue for pilots as in our study, but also in other domains, such as in healthcare~\cite{wang_designing_2019,jacobs_how_2021,fogliato_who_2022}. Process-oriented support in contrast keeps users engaged in their decision-making process and helps them to generate a decision themselves while reasoning forward. Recommendations are optional, serving as confirmation or challenge toward the end of the process rather than the starting point. As reflected by our participants' discussions, the core challenge for process-oriented support is to strike a good balance between interpreting information with AI and leaving the interpretation work to the human, which is strongly use-case-dependent. Our continuous support concept leaves much of the interpretation to pilots, which worked well for the diversion use case since pilots are highly trained expert users who usually know what information is important, as demonstrated by participants in the Baseline condition. For such expert users, a restrained form of AI support like continuous support has the benefit that it allows them to work very flexibly with the system, as discussed by our participants (\autoref{sec:tension}). Use cases with less expert users may require a ``stronger'' role for the AI, e.g., guiding users toward important information through explanations. Even for our use case, some pilots wished for more sophisticated AI functionality.

Second, the role of transparency changes. In recommendation-centric support, transparency aids appropriate reliance on recommendations. In process-oriented support, transparency facilitates appropriation~\cite{dix_designing_2007} by helping users understand how well the system functionality aligns with their momentary intentions and needs. The importance of appropriation in AI-assisted decision-making has been discussed by several authors recently~\cite{zhang_resilience_2023,cabitza_need_2021,ehsan_expanding_2021}. It differs from reliance calibration in at least two aspects. One is that transparency for appropriation is targeting the more granular level of intermediary decision-making steps, rather than the end of the process. This leads to more nuanced behavior than a simple accept-reject dichotomy, like partial reliance and negotiation patterns, as observed by Sivaraman~et~al.~\cite{sivaraman_ignore_2023} and with our participants as well. The other difference is that appropriate reliance concerns the correctness of the AI, while appropriation is about the user's intention. As was evident from our participants' requests, this does not necessarily involve explaining how the algorithm works.

Lastly, while control has been studied in AI systems for instance in the context of interactive machine learning~\cite{kulesza_principles_2015,dudley_review_2018,smith-renner_no_2020}, there is usually no control in recommendation-centric decision support. In process-oriented support, control is an important complement to transparency to enable appropriation. Notably, different to interactive machine learning, control is not necessarily about feedback that the model should learn, but rather about steering the system according to users' momentary intentions and context, as exemplified by the criteria editing feature in our system, or the controls designed by Cai~et~al. in a healthcare application~\cite{cai_human-centered_2019}.

Process-oriented support joins the ranks of recent alternatives to recommendation-centered support. In particular, we see it as a generalization of evaluative AI~\cite{miller_explainable_2023}. Core to process-oriented support is to identify the main challenges in a decision-making process which can be supported by AI. Evaluative AI can be interpreted as process-oriented support for decisions where the main challenge is to generate and evaluate multiple hypotheses, such as in medical diagnostics. In the diversion use case, the main challenge was different, namely the difficulty of information gathering. This leads to different support opportunities, such as our continuous support concept, which proved to be a good, but not perfect starting point. Further research is required to better understand how each element of process-oriented support can be implemented for diversion assistance and beyond, from useful AI support roles, over how to enable appropriation through transparency and control, to how recommendations can be integrated toward the end of users' decision-making process.

\subsection{Limitations and Future Work}
\label{sec:limitations}
There are several limitations to our results, mostly stemming from our flight simulation setup. For one, we decided to conduct the study with a single pilot at a time, rather than with two pilots as is the standard in today's cockpits. We are fully aware that this is a significant deviation from how pilots work today, as the cooperation of both pilots is fundamental to today's operations~\cite{helmreich_evolution_1999}. We still decided for a single-pilot setup for two reasons. First, recruiting pilots is challenging, and relying on the simultaneous availability of pairs of pilots would have significantly reduced our sample size or even made the study impossible to conduct. Second, we are interested in how different AI support paradigms would affect pilots' decision-making in a possible AI-augmented future of aviation. We are agnostic as to whether this future maintains a two-pilots cockpit or involves single-pilot operations, which is a long-standing goal of the aviation industry~\cite{harris_humancentred_2007}. 

Another limitation was that we could not provide all tools that pilots use in their daily work, such as maps or tools to calculate certain performance and flight data. Besides being prohibitively costly to replicate all these tools, they are also not standardized between airlines.
Still, participants were able to appropriate what was available to them in X-Plane to fulfill many of their information needs. Several participants also praised the study setup to be well executed. We therefore assume that the impact of the limited tools was not critical.

Furthermore, despite our best efforts to construct a representative range of valid scenarios, it was not possible in our setup to fully reflect complexities that can arise from the coordination with multiple stakeholders. 
Such coordination was rudimentarily present in Scenario 3, but in reality, these interactions would be much more open-ended and therefore hard to simulate in a controlled setting like ours. While studies like ours do produce valuable insights, other methods would be required to better capture the intricacies of such open-ended scenarios. For instance, one could analyze incident reports post deployment, which is a proven strategy in aviation to improve safety.

Lastly, given the difficulty and cost to recruit professional pilots, our sample size was relatively small for our quantitative analyses. We tried to address this limitation by triangulating the quantitative results with rich qualitative data from the think-aloud sessions and exit interviews. We found clear trends in the quantitative data that were consistent with the qualitative data.

Overall, these limitations highlight the difficulty of rigorously evaluating different support paradigms on a realistic task with domain experts. We suspect this is also the reason why such studies are rare. Nevertheless, similar to previous AI-assisted decision-making research on real-world tasks, our study produced insights that would likely not be available with simple tasks and crowd worker participants. We therefore encourage further research on realistic tasks with domain experts to close the gap between research and real-world adoption of AI decision support.

\section{Conclusion}
We conducted one of the first empirical comparisons between the dominant recommendation-centric AI decision support paradigm and alternative approaches in a realistic task setting with domain experts. We found that with recommendation-only, participants exhibited strong overreliance as they thought beyond the limits of the system significantly less than with the baseline variant. The recommendation-only variant was also surprisingly not faster than continuous support, in fact it even tended to be slower. We further found that to benefit from recommendations, pilots must engage with them while reasoning forward. Continuous support appeared to be an effective approach to encourage forward reasoning, allowing to display recommendations with less overreliance and leading to faster decisions when combined with recommendations. However, the effectiveness of continuous support was sensitive to disruptions between normal flight and abnormal situations. Our qualitative analysis reveals that pilots do not primarily value the system for suggesting suitable options, but rather for helping them to gain a quick overview of a large amount of information. While pilots welcome AI features for this purpose, they are concerned that too obtrusive AI takes away too much of the decision from them, with recommendations being particularly controversial. Participants' statements further revealed their requirements for transparency and control to appropriate the system. Notably, their primary concern with transparency was not how the AI works, but rather how it fits their intention. 

Our results challenge the assumption that AI decision support should be recommendation-centric and highlight the importance of supporting the decision-making process in a forward direction. Our continuous support concept was a promising first step in this direction, while our findings suggest many opportunities for improvement and future work. Further research is especially required to understand how to more robustly support decisions in a forward direction. We encapsulated our findings in a framework for process-oriented support, and envision our work as a contribution toward a more holistic perspective on AI-assisted decision-making that looks beyond recommendations and explanations.

\begin{acks}
This work was supported by the German Federal Ministry for Economic Affairs and Climate Action (BMWK) under the LuFo VI-1 program, project KIEZ4-0.

Many thanks to our participants, who were eager to discuss our research with us, were happy to provide us with valuable contacts and insights into their daily work, and spread the word about our study among their colleagues.
\end{acks}

\bibliographystyle{ACM-Reference-Format}
\newcommand{\shownote}[1]{\unskip}
\bibliography{references}


\begin{thebibliography}{76}


\ifx \showCODEN    \undefined \def \showCODEN     #1{\unskip}     \fi
\ifx \showDOI      \undefined \def \showDOI       #1{#1}\fi
\ifx \showISBNx    \undefined \def \showISBNx     #1{\unskip}     \fi
\ifx \showISBNxiii \undefined \def \showISBNxiii  #1{\unskip}     \fi
\ifx \showISSN     \undefined \def \showISSN      #1{\unskip}     \fi
\ifx \showLCCN     \undefined \def \showLCCN      #1{\unskip}     \fi
\ifx \shownote     \undefined \def \shownote      #1{#1}          \fi
\ifx \showarticletitle \undefined \def \showarticletitle #1{#1}   \fi
\ifx \showURL      \undefined \def \showURL       {\relax}        \fi
\providecommand\bibfield[2]{#2}
\providecommand\bibinfo[2]{#2}
\providecommand\natexlab[1]{#1}
\providecommand\showeprint[2][]{arXiv:#2}

\bibitem[Bach et~al\mbox{.}(2023)]%
        {bach_if_2023}
\bibfield{author}{\bibinfo{person}{Anne Kathrine~Petersen Bach}, \bibinfo{person}{Trine~Munch Nørgaard}, \bibinfo{person}{Jens~Christian Brok}, {and} \bibinfo{person}{Niels Van~Berkel}.} \bibinfo{year}{2023}\natexlab{}.
\newblock \showarticletitle{“{If} {I} had all the time in the world”: ophthalmologists’ perceptions of anchoring bias mitigation in clinical {AI} support}. In \bibinfo{booktitle}{\emph{Proceedings of the 2023 {CHI} {Conference} on {Human} {Factors} in {Computing} {Systems}}} \emph{(\bibinfo{series}{{CHI} '23})}. \bibinfo{publisher}{ACM}, \bibinfo{address}{Hamburg, Germany}, \bibinfo{pages}{16:1--16:14}.
\newblock
\showISBNx{978-1-4503-9421-5}
\urldef\tempurl%
\url{https://doi.org/10.1145/3544548.3581513}
\showDOI{\tempurl}


\bibitem[Bansal et~al\mbox{.}(2021)]%
        {bansal_does_2021}
\bibfield{author}{\bibinfo{person}{Gagan Bansal}, \bibinfo{person}{Tongshuang Wu}, \bibinfo{person}{Joyce Zhou}, \bibinfo{person}{Raymond Fok}, \bibinfo{person}{Besmira Nushi}, \bibinfo{person}{Ece Kamar}, \bibinfo{person}{Marco~Tulio Ribeiro}, {and} \bibinfo{person}{Daniel Weld}.} \bibinfo{year}{2021}\natexlab{}.
\newblock \showarticletitle{Does the whole exceed its parts? {The} effect of {AI} explanations on complementary team performance}. In \bibinfo{booktitle}{\emph{Proceedings of the 2021 {CHI} {Conference} on {Human} {Factors} in {Computing} {Systems}}} \emph{(\bibinfo{series}{{CHI} '21})}. \bibinfo{publisher}{ACM}, \bibinfo{address}{Yokohama, Japan}, \bibinfo{pages}{81:1--81:16}.
\newblock
\showISBNx{978-1-4503-8096-6}
\urldef\tempurl%
\url{https://doi.org/10.1145/3411764.3445717}
\showDOI{\tempurl}


\bibitem[Blandford et~al\mbox{.}(2016)]%
        {blandford_analysing_2016}
\bibfield{author}{\bibinfo{person}{Ann Blandford}, \bibinfo{person}{Dominic Furniss}, {and} \bibinfo{person}{Stephann Makri}.} \bibinfo{year}{2016}\natexlab{}.
\newblock \showarticletitle{Analysing {Data}}.
\newblock In \bibinfo{booktitle}{\emph{Qualitative {HCI} {Research}: {Going} {Behind} the {Scenes}}}, \bibfield{editor}{\bibinfo{person}{Ann Blandford}, \bibinfo{person}{Dominic Furniss}, {and} \bibinfo{person}{Stephann Makri}} (Eds.). \bibinfo{publisher}{Springer International Publishing}, \bibinfo{address}{Cham}, \bibinfo{pages}{51--60}.
\newblock
\showISBNx{978-3-031-02217-3}
\urldef\tempurl%
\url{https://doi.org/10.1007/978-3-031-02217-3_5}
\showDOI{\tempurl}


\bibitem[Blomberg et~al\mbox{.}(2018)]%
        {blomberg_acting_2018}
\bibfield{author}{\bibinfo{person}{Jeanette Blomberg}, \bibinfo{person}{Aly Megahed}, {and} \bibinfo{person}{Ray Strong}.} \bibinfo{year}{2018}\natexlab{}.
\newblock \showarticletitle{Acting on analytics: accuracy, precision, interpretation, and performativity}.
\newblock \bibinfo{journal}{\emph{Ethnographic Praxis in Industry Conference Proceedings}} \bibinfo{volume}{2018}, \bibinfo{number}{1} (\bibinfo{year}{2018}), \bibinfo{pages}{281--300}.
\newblock
\showISSN{1559-8918}
\urldef\tempurl%
\url{https://doi.org/10.1111/1559-8918.2018.01208}
\showDOI{\tempurl}


\bibitem[Burgess et~al\mbox{.}(2023)]%
        {burgess_healthcare_2023}
\bibfield{author}{\bibinfo{person}{Eleanor~R. Burgess}, \bibinfo{person}{Ivana Jankovic}, \bibinfo{person}{Melissa Austin}, \bibinfo{person}{Nancy Cai}, \bibinfo{person}{Adela Kapuścińska}, \bibinfo{person}{Suzanne Currie}, \bibinfo{person}{J.~Marc Overhage}, \bibinfo{person}{Erika~S Poole}, {and} \bibinfo{person}{Jofish Kaye}.} \bibinfo{year}{2023}\natexlab{}.
\newblock \showarticletitle{Healthcare {AI} treatment decision support: design principles to enhance clinician adoption and trust}. In \bibinfo{booktitle}{\emph{Proceedings of the 2023 {CHI} {Conference} on {Human} {Factors} in {Computing} {Systems}}} \emph{(\bibinfo{series}{{CHI} '23})}. \bibinfo{publisher}{ACM}, \bibinfo{address}{Hamburg, Germany}, \bibinfo{pages}{15:1--15:19}.
\newblock
\showISBNx{978-1-4503-9421-5}
\urldef\tempurl%
\url{https://doi.org/10.1145/3544548.3581251}
\showDOI{\tempurl}


\bibitem[Bussone et~al\mbox{.}(2015)]%
        {bussone_role_2015}
\bibfield{author}{\bibinfo{person}{Adrian Bussone}, \bibinfo{person}{Simone Stumpf}, {and} \bibinfo{person}{Dympna O'Sullivan}.} \bibinfo{year}{2015}\natexlab{}.
\newblock \showarticletitle{The role of explanations on trust and reliance in clinical decision support systems}. In \bibinfo{booktitle}{\emph{Proceedings of the 2015 {International} {Conference} on {Healthcare} {Informatics}}} \emph{(\bibinfo{series}{{ICHI} 2015})}. \bibinfo{publisher}{IEEE}, \bibinfo{address}{Dallas, TX, USA}, \bibinfo{pages}{160--169}.
\newblock
\showISBNx{978-1-4673-9548-9}
\urldef\tempurl%
\url{https://doi.org/10.1109/ICHI.2015.26}
\showDOI{\tempurl}


\bibitem[Buçinca et~al\mbox{.}(2022)]%
        {bucinca_beyond_2022}
\bibfield{author}{\bibinfo{person}{Zana Buçinca}, \bibinfo{person}{Alexandra Chouldechova}, \bibinfo{person}{Jennifer Wortman~Vaughan}, {and} \bibinfo{person}{Krzysztof~Z. Gajos}.} \bibinfo{year}{2022}\natexlab{}.
\newblock \showarticletitle{Beyond end predictions: stop putting machine learning first and design human-centered {AI} for decision support}. In \bibinfo{booktitle}{\emph{Virtual {Workshop} on {Human}-{Centered} {AI} {Workshop} at {NeurIPS} ({HCAI} @ {NeurIPS} '22)}}. \bibinfo{address}{Virtual Event, USA}, \bibinfo{pages}{1--4}.
\newblock


\bibitem[Buçinca et~al\mbox{.}(2021)]%
        {bucinca_trust_2021}
\bibfield{author}{\bibinfo{person}{Zana Buçinca}, \bibinfo{person}{Maja~Barbara Malaya}, {and} \bibinfo{person}{Krzysztof~Z. Gajos}.} \bibinfo{year}{2021}\natexlab{}.
\newblock \showarticletitle{To trust or to think: cognitive forcing functions can reduce overreliance on {AI} in {AI}-assisted decision-making}.
\newblock \bibinfo{journal}{\emph{Proceedings of the ACM on Human-Computer Interaction}} \bibinfo{volume}{5}, \bibinfo{number}{CSCW1} (\bibinfo{date}{April} \bibinfo{year}{2021}), \bibinfo{pages}{188:1--188:21}.
\newblock
\urldef\tempurl%
\url{https://doi.org/10.1145/3449287}
\showDOI{\tempurl}


\bibitem[Cabitza et~al\mbox{.}(2021)]%
        {cabitza_need_2021}
\bibfield{author}{\bibinfo{person}{Federico Cabitza}, \bibinfo{person}{Andrea Campagner}, {and} \bibinfo{person}{Carla Simone}.} \bibinfo{year}{2021}\natexlab{}.
\newblock \showarticletitle{The need to move away from agential-{AI}: empirical investigations, useful concepts and open issues}.
\newblock \bibinfo{journal}{\emph{International Journal of Human-Computer Studies}}  \bibinfo{volume}{155} (\bibinfo{date}{Nov.} \bibinfo{year}{2021}), \bibinfo{pages}{102696:1--102696:11}.
\newblock
\showISSN{10715819}
\urldef\tempurl%
\url{https://doi.org/10.1016/j.ijhcs.2021.102696}
\showDOI{\tempurl}


\bibitem[Cai et~al\mbox{.}(2019)]%
        {cai_human-centered_2019}
\bibfield{author}{\bibinfo{person}{Carrie~J. Cai}, \bibinfo{person}{Martin~C. Stumpe}, \bibinfo{person}{Michael Terry}, \bibinfo{person}{Emily Reif}, \bibinfo{person}{Narayan Hegde}, \bibinfo{person}{Jason Hipp}, \bibinfo{person}{Been Kim}, \bibinfo{person}{Daniel Smilkov}, \bibinfo{person}{Martin Wattenberg}, \bibinfo{person}{Fernanda Viegas}, {and} \bibinfo{person}{Greg~S. Corrado}.} \bibinfo{year}{2019}\natexlab{}.
\newblock \showarticletitle{Human-centered tools for coping with imperfect algorithms during medical decision-making}. In \bibinfo{booktitle}{\emph{Proceedings of the 2019 {CHI} {Conference} on {Human} {Factors} in {Computing} {Systems}}} \emph{(\bibinfo{series}{{CHI} '19})}. \bibinfo{publisher}{ACM}, \bibinfo{address}{Glasgow, Scotland, UK}, \bibinfo{pages}{4:1--4:14}.
\newblock
\showISBNx{978-1-4503-5970-2}
\urldef\tempurl%
\url{https://doi.org/10.1145/3290605.3300234}
\showDOI{\tempurl}


\bibitem[Cao(2022)]%
        {cao_ai_2022}
\bibfield{author}{\bibinfo{person}{Longbing Cao}.} \bibinfo{year}{2022}\natexlab{}.
\newblock \showarticletitle{{AI} in finance: challenges, techniques, and opportunities}.
\newblock \bibinfo{journal}{\emph{Comput. Surveys}} \bibinfo{volume}{55}, \bibinfo{number}{3} (\bibinfo{date}{Feb.} \bibinfo{year}{2022}), \bibinfo{pages}{64:1--64:38}.
\newblock
\showISSN{0360-0300}
\urldef\tempurl%
\url{https://doi.org/10.1145/3502289}
\showDOI{\tempurl}


\bibitem[Castelo et~al\mbox{.}(2019)]%
        {castelo_task-dependent_2019}
\bibfield{author}{\bibinfo{person}{Noah Castelo}, \bibinfo{person}{Maarten~W. Bos}, {and} \bibinfo{person}{Donald~R. Lehmann}.} \bibinfo{year}{2019}\natexlab{}.
\newblock \showarticletitle{Task-dependent algorithm aversion}.
\newblock \bibinfo{journal}{\emph{Journal of Marketing Research}} \bibinfo{volume}{56}, \bibinfo{number}{5} (\bibinfo{date}{Oct.} \bibinfo{year}{2019}), \bibinfo{pages}{809--825}.
\newblock
\showISSN{0022-2437}
\urldef\tempurl%
\url{https://doi.org/10.1177/0022243719851788}
\showDOI{\tempurl}
\newblock
\shownote{Publisher: SAGE Publications Inc}.


\bibitem[Cheng and Chouldechova(2023)]%
        {cheng_overcoming_2023}
\bibfield{author}{\bibinfo{person}{Lingwei Cheng} {and} \bibinfo{person}{Alexandra Chouldechova}.} \bibinfo{year}{2023}\natexlab{}.
\newblock \showarticletitle{Overcoming algorithm aversion: a comparison between process and outcome control}. In \bibinfo{booktitle}{\emph{Proceedings of the 2023 {CHI} {Conference} on {Human} {Factors} in {Computing} {Systems}}} \emph{(\bibinfo{series}{{CHI} '23})}. \bibinfo{publisher}{ACM}, \bibinfo{address}{Hamburg, Germany}, \bibinfo{pages}{756:1--756:27}.
\newblock
\showISBNx{978-1-4503-9421-5}
\urldef\tempurl%
\url{https://doi.org/10.1145/3544548.3581253}
\showDOI{\tempurl}


\bibitem[Dietvorst et~al\mbox{.}(2015)]%
        {dietvorst_algorithm_2015}
\bibfield{author}{\bibinfo{person}{Berkeley~J. Dietvorst}, \bibinfo{person}{Joseph~P. Simmons}, {and} \bibinfo{person}{Cade Massey}.} \bibinfo{year}{2015}\natexlab{}.
\newblock \showarticletitle{Algorithm aversion: people erroneously avoid algorithms after seeing them err}.
\newblock \bibinfo{journal}{\emph{Journal of Experimental Psychology: General}} \bibinfo{volume}{144}, \bibinfo{number}{1} (\bibinfo{year}{2015}), \bibinfo{pages}{114--126}.
\newblock
\showISSN{1939-2222, 0096-3445}
\urldef\tempurl%
\url{https://doi.org/10.1037/xge0000033}
\showDOI{\tempurl}


\bibitem[Dix(2007)]%
        {dix_designing_2007}
\bibfield{author}{\bibinfo{person}{Alan Dix}.} \bibinfo{year}{2007}\natexlab{}.
\newblock \showarticletitle{Designing for appropriation}. In \bibinfo{booktitle}{\emph{Proceedings of the 21st {British} {HCI} {Group} {Annual} {Conference} on {People} and {Computers}}} \emph{(\bibinfo{series}{{BCS}-{HCI} '07}, Vol.~\bibinfo{volume}{2})}. \bibinfo{publisher}{BCS Learning \& Development Ltd.}, \bibinfo{address}{Lancaster, UK}, \bibinfo{pages}{27--30}.
\newblock
\showISBNx{978-1-902505-95-4}
\urldef\tempurl%
\url{https://doi.org/10.14236/ewic/HCI2007.53}
\showDOI{\tempurl}


\bibitem[Dudley and Kristensson(2018)]%
        {dudley_review_2018}
\bibfield{author}{\bibinfo{person}{John~J. Dudley} {and} \bibinfo{person}{Per~Ola Kristensson}.} \bibinfo{year}{2018}\natexlab{}.
\newblock \showarticletitle{A review of user interface design for interactive machine learning}.
\newblock \bibinfo{journal}{\emph{ACM Transactions on Interactive Intelligent Systems}} \bibinfo{volume}{8}, \bibinfo{number}{2} (\bibinfo{date}{July} \bibinfo{year}{2018}), \bibinfo{pages}{8:1--8:37}.
\newblock
\showISSN{2160-6455, 2160-6463}
\urldef\tempurl%
\url{https://doi.org/10.1145/3185517}
\showDOI{\tempurl}


\bibitem[EASA(2023)]%
        {easa_artificial_2023}
\bibfield{author}{\bibinfo{person}{EASA}.} \bibinfo{year}{2023}\natexlab{}.
\newblock \bibinfo{booktitle}{\emph{Artificial {Intelligence} {Roadmap} 2.0: {A} human-centric approach to {AI} in aviation}}.
\newblock \bibinfo{type}{{T}echnical {R}eport}. \bibinfo{institution}{European Union Aviation Safety Agency (EASA)}. \bibinfo{pages}{36} pages.
\newblock
\urldef\tempurl%
\url{https://www.easa.europa.eu/en/document-library/general-publications/easa-artificial-intelligence-roadmap-20}
\showURL{%
\tempurl}


\bibitem[Ehsan et~al\mbox{.}(2021)]%
        {ehsan_expanding_2021}
\bibfield{author}{\bibinfo{person}{Upol Ehsan}, \bibinfo{person}{Q.~Vera Liao}, \bibinfo{person}{Michael Muller}, \bibinfo{person}{Mark~O. Riedl}, {and} \bibinfo{person}{Justin~D. Weisz}.} \bibinfo{year}{2021}\natexlab{}.
\newblock \showarticletitle{Expanding explainability: towards social transparency in {AI} systems}. In \bibinfo{booktitle}{\emph{Proceedings of the 2021 {CHI} {Conference} on {Human} {Factors} in {Computing} {Systems}}} \emph{(\bibinfo{series}{{CHI} '21})}. \bibinfo{publisher}{ACM}, \bibinfo{address}{Yokohama, Japan}, \bibinfo{pages}{82:1--82:19}.
\newblock
\showISBNx{978-1-4503-8096-6}
\urldef\tempurl%
\url{https://doi.org/10.1145/3411764.3445188}
\showDOI{\tempurl}


\bibitem[Eiband et~al\mbox{.}(2019)]%
        {eiband_impact_2019}
\bibfield{author}{\bibinfo{person}{Malin Eiband}, \bibinfo{person}{Daniel Buschek}, \bibinfo{person}{Alexander Kremer}, {and} \bibinfo{person}{Heinrich Hussmann}.} \bibinfo{year}{2019}\natexlab{}.
\newblock \showarticletitle{The impact of placebic explanations on trust in intelligent systems}. In \bibinfo{booktitle}{\emph{Extended {Abstracts} of the 2019 {CHI} {Conference} on {Human} {Factors} in {Computing} {Systems}}} \emph{(\bibinfo{series}{{CHI} {EA} '19})}. \bibinfo{publisher}{ACM}, \bibinfo{address}{Glasgow, Scotland, UK}, \bibinfo{pages}{LBW0243:1--LBW0243:6}.
\newblock
\showISBNx{978-1-4503-5971-9}
\urldef\tempurl%
\url{https://doi.org/10.1145/3290607.3312787}
\showDOI{\tempurl}


\bibitem[Feger et~al\mbox{.}(2022)]%
        {feger_hci_2022}
\bibfield{author}{\bibinfo{person}{Sebastian~S. Feger}, \bibinfo{person}{Felix Ehrentraut}, \bibinfo{person}{Christopher Katins}, \bibinfo{person}{Philippe Palanque}, {and} \bibinfo{person}{Thomas Kosch}.} \bibinfo{year}{2022}\natexlab{}.
\newblock \showarticletitle{{HCI} for general aviation: current state and research challenges}.
\newblock \bibinfo{journal}{\emph{Interactions}} \bibinfo{volume}{29}, \bibinfo{number}{6} (\bibinfo{date}{Nov.} \bibinfo{year}{2022}), \bibinfo{pages}{60--65}.
\newblock
\showISSN{1072-5520}
\urldef\tempurl%
\url{https://doi.org/10.1145/3564040}
\showDOI{\tempurl}


\bibitem[Fogliato et~al\mbox{.}(2022)]%
        {fogliato_who_2022}
\bibfield{author}{\bibinfo{person}{Riccardo Fogliato}, \bibinfo{person}{Shreya Chappidi}, \bibinfo{person}{Matthew Lungren}, \bibinfo{person}{Paul Fisher}, \bibinfo{person}{Diane Wilson}, \bibinfo{person}{Michael Fitzke}, \bibinfo{person}{Mark Parkinson}, \bibinfo{person}{Eric Horvitz}, \bibinfo{person}{Kori Inkpen}, {and} \bibinfo{person}{Besmira Nushi}.} \bibinfo{year}{2022}\natexlab{}.
\newblock \showarticletitle{Who goes first? {Influences} of human-{AI} workflow on decision making in clinical imaging}. In \bibinfo{booktitle}{\emph{Proceedings of the 2022 {ACM} {Conference} on {Fairness}, {Accountability}, and {Transparency}}} \emph{(\bibinfo{series}{{FAccT} '22})}. \bibinfo{publisher}{Association for Computing Machinery}, \bibinfo{address}{Seoul, Republic of Korea}, \bibinfo{pages}{1362--1374}.
\newblock
\showISBNx{978-1-4503-9352-2}
\urldef\tempurl%
\url{https://doi.org/10.1145/3531146.3533193}
\showDOI{\tempurl}


\bibitem[Fok and Weld(2024)]%
        {fok_search_2024}
\bibfield{author}{\bibinfo{person}{Raymond Fok} {and} \bibinfo{person}{Daniel~S. Weld}.} \bibinfo{year}{2024}\natexlab{}.
\newblock \showarticletitle{In search of verifiability: {Explanations} rarely enable complementary performance in {AI}-advised decision making}.
\newblock \bibinfo{journal}{\emph{AI Magazine}} \bibinfo{number}{Early View} (\bibinfo{date}{July} \bibinfo{year}{2024}), \bibinfo{pages}{1--16}.
\newblock
\showISSN{2371-9621}
\urldef\tempurl%
\url{https://doi.org/10.1002/aaai.12182}
\showDOI{\tempurl}


\bibitem[Gajos and Mamykina(2022)]%
        {gajos_people_2022}
\bibfield{author}{\bibinfo{person}{Krzysztof~Z. Gajos} {and} \bibinfo{person}{Lena Mamykina}.} \bibinfo{year}{2022}\natexlab{}.
\newblock \showarticletitle{Do people engage cognitively with {AI}? {Impact} of {AI} assistance on incidental learning}. In \bibinfo{booktitle}{\emph{27th {International} {Conference} on {Intelligent} {User} {Interfaces}}} \emph{(\bibinfo{series}{{IUI} '22})}. \bibinfo{publisher}{ACM}, \bibinfo{address}{Helsinki, Finland}, \bibinfo{pages}{794--806}.
\newblock
\showISBNx{978-1-4503-9144-3}
\urldef\tempurl%
\url{https://doi.org/10.1145/3490099.3511138}
\showDOI{\tempurl}


\bibitem[Green and Chen(2019)]%
        {green_principles_2019}
\bibfield{author}{\bibinfo{person}{Ben Green} {and} \bibinfo{person}{Yiling Chen}.} \bibinfo{year}{2019}\natexlab{}.
\newblock \showarticletitle{The principles and limits of algorithm-in-the-loop decision making}.
\newblock \bibinfo{journal}{\emph{Proceedings of the ACM on Human-Computer Interaction}} \bibinfo{volume}{3}, \bibinfo{number}{CSCW} (\bibinfo{date}{Nov.} \bibinfo{year}{2019}), \bibinfo{pages}{50:1--50:24}.
\newblock
\urldef\tempurl%
\url{https://doi.org/10.1145/3359152}
\showDOI{\tempurl}


\bibitem[Gu et~al\mbox{.}(2023)]%
        {gu_augmenting_2023}
\bibfield{author}{\bibinfo{person}{Hongyan Gu}, \bibinfo{person}{Chunxu Yang}, \bibinfo{person}{Mohammad Haeri}, \bibinfo{person}{Jing Wang}, \bibinfo{person}{Shirley Tang}, \bibinfo{person}{Wenzhong Yan}, \bibinfo{person}{Shujin He}, \bibinfo{person}{Christopher~Kazu Williams}, \bibinfo{person}{Shino Magaki}, {and} \bibinfo{person}{Xiang~'Anthony' Chen}.} \bibinfo{year}{2023}\natexlab{}.
\newblock \showarticletitle{Augmenting pathologists with {NaviPath}: design and evaluation of a human-{AI} collaborative navigation system}. In \bibinfo{booktitle}{\emph{Proceedings of the 2023 {CHI} {Conference} on {Human} {Factors} in {Computing} {Systems}}} \emph{(\bibinfo{series}{{CHI} '23})}. \bibinfo{publisher}{ACM}, \bibinfo{address}{Hamburg, Germany}, \bibinfo{pages}{349:1--349:19}.
\newblock
\showISBNx{978-1-4503-9421-5}
\urldef\tempurl%
\url{https://doi.org/10.1145/3544548.3580694}
\showDOI{\tempurl}


\bibitem[Harris(2007)]%
        {harris_humancentred_2007}
\bibfield{author}{\bibinfo{person}{Don Harris}.} \bibinfo{year}{2007}\natexlab{}.
\newblock \showarticletitle{A human‐centred design agenda for the development of single crew operated commercial aircraft}.
\newblock \bibinfo{journal}{\emph{Aircraft Engineering and Aerospace Technology}} \bibinfo{volume}{79}, \bibinfo{number}{5} (\bibinfo{date}{Sept.} \bibinfo{year}{2007}), \bibinfo{pages}{518--526}.
\newblock
\showISSN{0002-2667}
\urldef\tempurl%
\url{https://doi.org/10.1108/00022660710780650}
\showDOI{\tempurl}


\bibitem[He et~al\mbox{.}(2023)]%
        {he_how_2023}
\bibfield{author}{\bibinfo{person}{Gaole He}, \bibinfo{person}{Stefan Buijsman}, {and} \bibinfo{person}{Ujwal Gadiraju}.} \bibinfo{year}{2023}\natexlab{}.
\newblock \showarticletitle{How stated accuracy of an {AI} system and analogies to explain accuracy affect human reliance on the system}.
\newblock \bibinfo{journal}{\emph{Proceedings of the ACM on Human-Computer Interaction}} \bibinfo{volume}{7}, \bibinfo{number}{CSCW2} (\bibinfo{date}{Oct.} \bibinfo{year}{2023}), \bibinfo{pages}{276:1--276:29}.
\newblock
\urldef\tempurl%
\url{https://doi.org/10.1145/3610067}
\showDOI{\tempurl}


\bibitem[Helmreich et~al\mbox{.}(1999)]%
        {helmreich_evolution_1999}
\bibfield{author}{\bibinfo{person}{Robert~L. Helmreich}, \bibinfo{person}{Ashleigh~C. Merritt}, {and} \bibinfo{person}{John~A. Wilhelm}.} \bibinfo{year}{1999}\natexlab{}.
\newblock \showarticletitle{The evolution of crew resource management training in commercial aviation}.
\newblock \bibinfo{journal}{\emph{The International Journal of Aviation Psychology}} \bibinfo{volume}{9}, \bibinfo{number}{1} (\bibinfo{year}{1999}), \bibinfo{pages}{19--32}.
\newblock
\urldef\tempurl%
\url{https://doi.org/10.1207/s15327108ijap0901_2}
\showDOI{\tempurl}


\bibitem[Hemmer et~al\mbox{.}(2023)]%
        {hemmer_human-ai_2023}
\bibfield{author}{\bibinfo{person}{Patrick Hemmer}, \bibinfo{person}{Monika Westphal}, \bibinfo{person}{Max Schemmer}, \bibinfo{person}{Sebastian Vetter}, \bibinfo{person}{Michael Vössing}, {and} \bibinfo{person}{Gerhard Satzger}.} \bibinfo{year}{2023}\natexlab{}.
\newblock \showarticletitle{Human-{AI} collaboration: the effect of {AI} delegation on human task performance and task satisfaction}. In \bibinfo{booktitle}{\emph{Proceedings of the 28th {International} {Conference} on {Intelligent} {User} {Interfaces}}} \emph{(\bibinfo{series}{{IUI} '23})}. \bibinfo{publisher}{ACM}, \bibinfo{address}{Sydney, NSW, Australia}, \bibinfo{pages}{453--463}.
\newblock
\showISBNx{9798400701061}
\urldef\tempurl%
\url{https://doi.org/10.1145/3581641.3584052}
\showDOI{\tempurl}


\bibitem[Hörmann(1994)]%
        {hormann_for-dec_1994}
\bibfield{author}{\bibinfo{person}{Hans-Jürgen Hörmann}.} \bibinfo{year}{1994}\natexlab{}.
\newblock \showarticletitle{{FOR}-{DEC} - {A} prescriptive model for aeronautical decision making}. In \bibinfo{booktitle}{\emph{Proceedings of the 21st {Conference} of the {European} {Association} for {Aviation} {Psychology} ({EAAP})}}. \bibinfo{publisher}{Avebury Aviation}, \bibinfo{address}{Dublin, Ireland}, \bibinfo{pages}{17--23}.
\newblock


\bibitem[Jacobs et~al\mbox{.}(2021a)]%
        {jacobs_designing_2021}
\bibfield{author}{\bibinfo{person}{Maia Jacobs}, \bibinfo{person}{Jeffrey He}, \bibinfo{person}{Melanie~F Pradier}, \bibinfo{person}{Barbara Lam}, \bibinfo{person}{Andrew~C Ahn}, \bibinfo{person}{Thomas~H McCoy}, \bibinfo{person}{Roy~H Perlis}, \bibinfo{person}{Finale Doshi-Velez}, {and} \bibinfo{person}{Krzysztof~Z Gajos}.} \bibinfo{year}{2021}\natexlab{a}.
\newblock \showarticletitle{Designing {AI} for trust and collaboration in time-constrained medical decisions: a sociotechnical lens}. In \bibinfo{booktitle}{\emph{Proceedings of the 2021 {CHI} {Conference} on {Human} {Factors} in {Computing} {Systems}}} \emph{(\bibinfo{series}{{CHI} '21})}. \bibinfo{publisher}{ACM}, \bibinfo{address}{Yokohama, Japan}, \bibinfo{pages}{659:1--659:14}.
\newblock
\urldef\tempurl%
\url{https://doi.org/10.1145/3411764.3445385}
\showDOI{\tempurl}


\bibitem[Jacobs et~al\mbox{.}(2021b)]%
        {jacobs_how_2021}
\bibfield{author}{\bibinfo{person}{Maia Jacobs}, \bibinfo{person}{Melanie~F. Pradier}, \bibinfo{person}{Thomas~H. McCoy}, \bibinfo{person}{Roy~H. Perlis}, \bibinfo{person}{Finale Doshi-Velez}, {and} \bibinfo{person}{Krzysztof~Z. Gajos}.} \bibinfo{year}{2021}\natexlab{b}.
\newblock \showarticletitle{How machine-learning recommendations influence clinician treatment selections: the example of antidepressant selection}.
\newblock \bibinfo{journal}{\emph{Translational Psychiatry}} \bibinfo{volume}{11}, \bibinfo{number}{1} (\bibinfo{date}{June} \bibinfo{year}{2021}), \bibinfo{pages}{108:1--108:9}.
\newblock
\showISSN{2158-3188}
\urldef\tempurl%
\url{https://doi.org/10.1038/s41398-021-01224-x}
\showDOI{\tempurl}


\bibitem[Kaltenhauser et~al\mbox{.}(2020)]%
        {kaltenhauser_you_2020}
\bibfield{author}{\bibinfo{person}{Annika Kaltenhauser}, \bibinfo{person}{Verena Rheinstädter}, \bibinfo{person}{Andreas Butz}, {and} \bibinfo{person}{Dieter~P. Wallach}.} \bibinfo{year}{2020}\natexlab{}.
\newblock \showarticletitle{"{You} have to piece the puzzle together": implications for designing decision support in intensive care}. In \bibinfo{booktitle}{\emph{Proceedings of the 2020 {ACM} {Designing} {Interactive} {Systems} {Conference}}} \emph{(\bibinfo{series}{{DIS} '20})}. \bibinfo{publisher}{ACM}, \bibinfo{address}{Eindhoven, Netherlands}, \bibinfo{pages}{1509--1522}.
\newblock
\showISBNx{978-1-4503-6974-9}
\urldef\tempurl%
\url{https://doi.org/10.1145/3357236.3395436}
\showDOI{\tempurl}


\bibitem[Kawakami et~al\mbox{.}(2022a)]%
        {kawakami_improving_2022}
\bibfield{author}{\bibinfo{person}{Anna Kawakami}, \bibinfo{person}{Venkatesh Sivaraman}, \bibinfo{person}{Hao-Fei Cheng}, \bibinfo{person}{Logan Stapleton}, \bibinfo{person}{Yanghuidi Cheng}, \bibinfo{person}{Diana Qing}, \bibinfo{person}{Adam Perer}, \bibinfo{person}{Zhiwei~Steven Wu}, \bibinfo{person}{Haiyi Zhu}, {and} \bibinfo{person}{Kenneth Holstein}.} \bibinfo{year}{2022}\natexlab{a}.
\newblock \showarticletitle{Improving human-{AI} partnerships in child welfare: understanding worker practices, challenges, and desires for algorithmic decision support}. In \bibinfo{booktitle}{\emph{Proceedings of the 2022 {CHI} {Conference} on {Human} {Factors} in {Computing} {Systems}}} \emph{(\bibinfo{series}{{CHI} '22})}. \bibinfo{publisher}{ACM}, \bibinfo{address}{New Orleans, LA, USA}, \bibinfo{pages}{52:1--52:18}.
\newblock
\showISBNx{978-1-4503-9157-3}
\urldef\tempurl%
\url{https://doi.org/10.1145/3491102.3517439}
\showDOI{\tempurl}


\bibitem[Kawakami et~al\mbox{.}(2022b)]%
        {kawakami_why_2022}
\bibfield{author}{\bibinfo{person}{Anna Kawakami}, \bibinfo{person}{Venkatesh Sivaraman}, \bibinfo{person}{Logan Stapleton}, \bibinfo{person}{Hao-Fei Cheng}, \bibinfo{person}{Adam Perer}, \bibinfo{person}{Zhiwei~Steven Wu}, \bibinfo{person}{Haiyi Zhu}, {and} \bibinfo{person}{Kenneth Holstein}.} \bibinfo{year}{2022}\natexlab{b}.
\newblock \showarticletitle{“{Why} do {I} care what’s similar?” {Probing} challenges in {AI}-assisted child welfare decision-making through worker-{AI} interface design concepts}. In \bibinfo{booktitle}{\emph{Designing {Interactive} {Systems} {Conference}}} \emph{(\bibinfo{series}{{DIS} '22})}. \bibinfo{publisher}{ACM}, \bibinfo{address}{Virtual Event, Australia}, \bibinfo{pages}{454--470}.
\newblock
\showISBNx{978-1-4503-9358-4}
\urldef\tempurl%
\url{https://doi.org/10.1145/3532106.3533556}
\showDOI{\tempurl}


\bibitem[Koon(2022)]%
        {koon_human-capabilities_2022}
\bibfield{author}{\bibinfo{person}{Sean Koon}.} \bibinfo{year}{2022}\natexlab{}.
\newblock \showarticletitle{A human-capabilities orientation for human-{AI} interaction design}. In \bibinfo{booktitle}{\emph{Virtual {Workshop} on {Human}-{Centered} {AI} {Workshop} at {NeurIPS} ({HCAI} @ {NeurIPS} '22)}}. \bibinfo{address}{Virtual Event, USA}, \bibinfo{pages}{1--5}.
\newblock


\bibitem[Kulesza et~al\mbox{.}(2015)]%
        {kulesza_principles_2015}
\bibfield{author}{\bibinfo{person}{Todd Kulesza}, \bibinfo{person}{Margaret Burnett}, \bibinfo{person}{Weng-Keen Wong}, {and} \bibinfo{person}{Simone Stumpf}.} \bibinfo{year}{2015}\natexlab{}.
\newblock \showarticletitle{Principles of explanatory debugging to personalize interactive machine learning}. In \bibinfo{booktitle}{\emph{Proceedings of the 20th {International} {Conference} on {Intelligent} {User} {Interfaces}}} \emph{(\bibinfo{series}{{IUI} '15})}. \bibinfo{publisher}{Association for Computing Machinery}, \bibinfo{address}{Atlanta, GA, USA}, \bibinfo{pages}{126--137}.
\newblock
\showISBNx{978-1-4503-3306-1}
\urldef\tempurl%
\url{https://doi.org/10.1145/2678025.2701399}
\showDOI{\tempurl}


\bibitem[Lai et~al\mbox{.}(2023)]%
        {lai_towards_2023}
\bibfield{author}{\bibinfo{person}{Vivian Lai}, \bibinfo{person}{Chacha Chen}, \bibinfo{person}{Alison Smith-Renner}, \bibinfo{person}{Q.~Vera Liao}, {and} \bibinfo{person}{Chenhao Tan}.} \bibinfo{year}{2023}\natexlab{}.
\newblock \showarticletitle{Towards a science of human-{AI} decision making: an overview of design space in empirical human-subject studies}. In \bibinfo{booktitle}{\emph{Proceedings of the 2023 {ACM} {Conference} on {Fairness}, {Accountability}, and {Transparency}}} \emph{(\bibinfo{series}{{FAccT} '23})}. \bibinfo{publisher}{ACM}, \bibinfo{address}{Chicago, IL, USA}, \bibinfo{pages}{1369--1385}.
\newblock
\showISBNx{9798400701924}
\urldef\tempurl%
\url{https://doi.org/10.1145/3593013.3594087}
\showDOI{\tempurl}


\bibitem[Lai and Tan(2019)]%
        {lai_human_2019}
\bibfield{author}{\bibinfo{person}{Vivian Lai} {and} \bibinfo{person}{Chenhao Tan}.} \bibinfo{year}{2019}\natexlab{}.
\newblock \showarticletitle{On human predictions with explanations and predictions of machine learning models: a case study on deception detection}. In \bibinfo{booktitle}{\emph{Proceedings of the {Conference} on {Fairness}, {Accountability}, and {Transparency}}} \emph{(\bibinfo{series}{{FAT}* '19})}. \bibinfo{publisher}{ACM}, \bibinfo{address}{Atlanta, GA, USA}, \bibinfo{pages}{29--38}.
\newblock
\showISBNx{978-1-4503-6125-5}
\urldef\tempurl%
\url{https://doi.org/10.1145/3287560.3287590}
\showDOI{\tempurl}


\bibitem[Lewis et~al\mbox{.}(2021)]%
        {lewis_data-driven_2021}
\bibfield{author}{\bibinfo{person}{Bridget~A. Lewis}, \bibinfo{person}{Valerie~J. Gawron}, \bibinfo{person}{Ehsan Esmaeilzadeh}, \bibinfo{person}{Ralf~H. Mayer}, \bibinfo{person}{Felipe Moreno-Hines}, \bibinfo{person}{Neil Nerwich}, {and} \bibinfo{person}{Paulo~M. Alves}.} \bibinfo{year}{2021}\natexlab{}.
\newblock \showarticletitle{Data-driven estimation of the impact of diversions due to in-flight medical emergencies on flight delay and aircraft operating costs}.
\newblock \bibinfo{journal}{\emph{Aerospace Medicine and Human Performance}} \bibinfo{volume}{92}, \bibinfo{number}{2} (\bibinfo{date}{Feb.} \bibinfo{year}{2021}), \bibinfo{pages}{99--105}.
\newblock
\urldef\tempurl%
\url{https://doi.org/10.3357/AMHP.5720.2021}
\showDOI{\tempurl}


\bibitem[Lim et~al\mbox{.}(2023)]%
        {lim_diagrammatization_2023}
\bibfield{author}{\bibinfo{person}{Brian~Y. Lim}, \bibinfo{person}{Joseph~P. Cahaly}, \bibinfo{person}{Chester Y.~F. Sng}, {and} \bibinfo{person}{Adam Chew}.} \bibinfo{year}{2023}\natexlab{}.
\newblock \bibinfo{title}{Diagrammatization: {Rationalizing} with diagrammatic {AI} explanations for abductive-deductive reasoning on hypotheses}.
\newblock
\newblock
\urldef\tempurl%
\url{https://doi.org/10.48550/arXiv.2302.01241}
\showDOI{\tempurl}
\newblock
\shownote{arXiv:2302.01241 [cs]}.


\bibitem[Lindvall et~al\mbox{.}(2021)]%
        {lindvall_rapid_2021}
\bibfield{author}{\bibinfo{person}{Martin Lindvall}, \bibinfo{person}{Claes Lundström}, {and} \bibinfo{person}{Jonas Löwgren}.} \bibinfo{year}{2021}\natexlab{}.
\newblock \showarticletitle{Rapid assisted visual search: supporting digital pathologists with imperfect {AI}}. In \bibinfo{booktitle}{\emph{Proceedings of the 26th {International} {Conference} on {Intelligent} {User} {Interfaces}}} \emph{(\bibinfo{series}{{IUI} '21})}. \bibinfo{publisher}{ACM}, \bibinfo{address}{College Station, TX, USA}, \bibinfo{pages}{504--513}.
\newblock
\showISBNx{978-1-4503-8017-1}
\urldef\tempurl%
\url{https://doi.org/10.1145/3397481.3450681}
\showDOI{\tempurl}


\bibitem[Liu et~al\mbox{.}(2021)]%
        {liu_understanding_2021}
\bibfield{author}{\bibinfo{person}{Han Liu}, \bibinfo{person}{Vivian Lai}, {and} \bibinfo{person}{Chenhao Tan}.} \bibinfo{year}{2021}\natexlab{}.
\newblock \showarticletitle{Understanding the effect of out-of-distribution examples and interactive explanations on human-{AI} decision making}.
\newblock \bibinfo{journal}{\emph{Proceedings of the ACM on Human-Computer Interaction}} \bibinfo{volume}{5}, \bibinfo{number}{CSCW2} (\bibinfo{date}{Oct.} \bibinfo{year}{2021}), \bibinfo{pages}{408:1--408:45}.
\newblock
\showISSN{2573-0142}
\urldef\tempurl%
\url{https://doi.org/10.1145/3479552}
\showDOI{\tempurl}


\bibitem[McGuirl and Sarter(2006)]%
        {mcguirl_supporting_2006}
\bibfield{author}{\bibinfo{person}{John~M. McGuirl} {and} \bibinfo{person}{Nadine~B. Sarter}.} \bibinfo{year}{2006}\natexlab{}.
\newblock \showarticletitle{Supporting trust calibration and the effective use of decision aids by presenting dynamic system confidence information}.
\newblock \bibinfo{journal}{\emph{Human Factors: The Journal of the Human Factors and Ergonomics Society}} \bibinfo{volume}{48}, \bibinfo{number}{4} (\bibinfo{date}{Dec.} \bibinfo{year}{2006}), \bibinfo{pages}{656--665}.
\newblock
\showISSN{0018-7208, 1547-8181}
\urldef\tempurl%
\url{https://doi.org/10.1518/001872006779166334}
\showDOI{\tempurl}


\bibitem[Miller(2023)]%
        {miller_explainable_2023}
\bibfield{author}{\bibinfo{person}{Tim Miller}.} \bibinfo{year}{2023}\natexlab{}.
\newblock \showarticletitle{Explainable {AI} is dead, long live explainable {AI}! {Hypothesis}-driven decision support using evaluative {AI}}. In \bibinfo{booktitle}{\emph{Proceedings of the 2023 {ACM} {Conference} on {Fairness}, {Accountability}, and {Transparency}}} \emph{(\bibinfo{series}{{FAccT} '23})}. \bibinfo{publisher}{ACM}, \bibinfo{address}{Chicago, IL, USA}, \bibinfo{pages}{333--342}.
\newblock
\showISBNx{9798400701924}
\urldef\tempurl%
\url{https://doi.org/10.1145/3593013.3594001}
\showDOI{\tempurl}


\bibitem[Park et~al\mbox{.}(2019)]%
        {park_slow_2019}
\bibfield{author}{\bibinfo{person}{Joon~Sung Park}, \bibinfo{person}{Rick Barber}, \bibinfo{person}{Alex Kirlik}, {and} \bibinfo{person}{Karrie Karahalios}.} \bibinfo{year}{2019}\natexlab{}.
\newblock \showarticletitle{A slow algorithm improves users' assessments of the algorithm's accuracy}.
\newblock \bibinfo{journal}{\emph{Proceedings of the ACM on Human-Computer Interaction}} \bibinfo{volume}{3}, \bibinfo{number}{CSCW} (\bibinfo{date}{Nov.} \bibinfo{year}{2019}), \bibinfo{pages}{102:1--102:15}.
\newblock
\urldef\tempurl%
\url{https://doi.org/10.1145/3359204}
\showDOI{\tempurl}


\bibitem[Poursabzi-Sangdeh et~al\mbox{.}(2021)]%
        {poursabzi-sangdeh_manipulating_2021}
\bibfield{author}{\bibinfo{person}{Forough Poursabzi-Sangdeh}, \bibinfo{person}{Daniel~G. Goldstein}, \bibinfo{person}{Jake~M. Hofman}, \bibinfo{person}{Jennifer~Wortman Vaughan}, {and} \bibinfo{person}{Hanna Wallach}.} \bibinfo{year}{2021}\natexlab{}.
\newblock \showarticletitle{Manipulating and measuring model interpretability}. In \bibinfo{booktitle}{\emph{Proceedings of the 2021 {CHI} {Conference} on {Human} {Factors} in {Computing} {Systems}}} \emph{(\bibinfo{series}{{CHI} '21})}. \bibinfo{publisher}{ACM}, \bibinfo{address}{Yokohama, Japan}, \bibinfo{pages}{237:1--237:52}.
\newblock
\showISBNx{978-1-4503-8096-6}
\urldef\tempurl%
\url{https://doi.org/10.1145/3411764.3445315}
\showDOI{\tempurl}


\bibitem[Prabhudesai et~al\mbox{.}(2023)]%
        {prabhudesai_understanding_2023}
\bibfield{author}{\bibinfo{person}{Snehal Prabhudesai}, \bibinfo{person}{Leyao Yang}, \bibinfo{person}{Sumit Asthana}, \bibinfo{person}{Xun Huan}, \bibinfo{person}{Q.~Vera Liao}, {and} \bibinfo{person}{Nikola Banovic}.} \bibinfo{year}{2023}\natexlab{}.
\newblock \showarticletitle{Understanding uncertainty: how lay decision-makers perceive and interpret uncertainty in human-{AI} decision making}. In \bibinfo{booktitle}{\emph{Proceedings of the 28th {International} {Conference} on {Intelligent} {User} {Interfaces}}} \emph{(\bibinfo{series}{{IUI} '23})}. \bibinfo{publisher}{ACM}, \bibinfo{address}{Sydney, NSW, Australia}, \bibinfo{pages}{379--396}.
\newblock
\showISBNx{9798400701061}
\urldef\tempurl%
\url{https://doi.org/10.1145/3581641.3584033}
\showDOI{\tempurl}


\bibitem[Prahl and Van~Swol(2017)]%
        {prahl_understanding_2017}
\bibfield{author}{\bibinfo{person}{Andrew Prahl} {and} \bibinfo{person}{Lyn Van~Swol}.} \bibinfo{year}{2017}\natexlab{}.
\newblock \showarticletitle{Understanding algorithm aversion: when is advice from automation discounted?}
\newblock \bibinfo{journal}{\emph{Journal of Forecasting}} \bibinfo{volume}{36}, \bibinfo{number}{6} (\bibinfo{year}{2017}), \bibinfo{pages}{691--702}.
\newblock
\showISSN{1099-131X}
\urldef\tempurl%
\url{https://doi.org/10.1002/for.2464}
\showDOI{\tempurl}


\bibitem[Raaijmakers(2019)]%
        {raaijmakers_artificial_2019}
\bibfield{author}{\bibinfo{person}{Stephan Raaijmakers}.} \bibinfo{year}{2019}\natexlab{}.
\newblock \showarticletitle{Artificial intelligence for law enforcement: challenges and opportunities}.
\newblock \bibinfo{journal}{\emph{IEEE Security \& Privacy}} \bibinfo{volume}{17}, \bibinfo{number}{5} (\bibinfo{date}{Sept.} \bibinfo{year}{2019}), \bibinfo{pages}{74--77}.
\newblock
\showISSN{1558-4046}
\urldef\tempurl%
\url{https://doi.org/10.1109/MSEC.2019.2925649}
\showDOI{\tempurl}


\bibitem[Rastogi et~al\mbox{.}(2022)]%
        {rastogi_deciding_2022}
\bibfield{author}{\bibinfo{person}{Charvi Rastogi}, \bibinfo{person}{Yunfeng Zhang}, \bibinfo{person}{Dennis Wei}, \bibinfo{person}{Kush~R. Varshney}, \bibinfo{person}{Amit Dhurandhar}, {and} \bibinfo{person}{Richard Tomsett}.} \bibinfo{year}{2022}\natexlab{}.
\newblock \showarticletitle{Deciding fast and slow: the role of cognitive biases in {AI}-assisted decision-making}.
\newblock \bibinfo{journal}{\emph{Proceedings of the ACM on Human-Computer Interaction}} \bibinfo{volume}{6}, \bibinfo{number}{CSCW1} (\bibinfo{date}{April} \bibinfo{year}{2022}), \bibinfo{pages}{83:1--83:22}.
\newblock
\urldef\tempurl%
\url{https://doi.org/10.1145/3512930}
\showDOI{\tempurl}


\bibitem[Rechkemmer and Yin(2022)]%
        {rechkemmer_when_2022}
\bibfield{author}{\bibinfo{person}{Amy Rechkemmer} {and} \bibinfo{person}{Ming Yin}.} \bibinfo{year}{2022}\natexlab{}.
\newblock \showarticletitle{When confidence meets accuracy: exploring the effects of multiple performance indicators on trust in machine learning models}. In \bibinfo{booktitle}{\emph{Proceedings of the 2022 {CHI} {Conference} on {Human} {Factors} in {Computing} {Systems}}} \emph{(\bibinfo{series}{{CHI} '22})}. \bibinfo{publisher}{Association for Computing Machinery}, \bibinfo{address}{New Orleans, LA, USA}, \bibinfo{pages}{535:1--535:14}.
\newblock
\showISBNx{978-1-4503-9157-3}
\urldef\tempurl%
\url{https://doi.org/10.1145/3491102.3501967}
\showDOI{\tempurl}


\bibitem[Roth et~al\mbox{.}(2022)]%
        {roth_methods_2022}
\bibfield{author}{\bibinfo{person}{Emilie Roth}, \bibinfo{person}{Devorah Klein}, \bibinfo{person}{Christen Sushereba}, \bibinfo{person}{Katie Ernst}, {and} \bibinfo{person}{Lauren Militello}.} \bibinfo{year}{2022}\natexlab{}.
\newblock \bibinfo{booktitle}{\emph{Methods and measures to evaluate technologies that influence aviator decision making and situation awareness}}.
\newblock \bibinfo{type}{Contract {Report}} USAARL-TECH-CR--2022-22. \bibinfo{institution}{Applied Decision Science}, \bibinfo{address}{Cincinnati, OH, USA}. \bibinfo{pages}{80} pages.
\newblock


\bibitem[Schemmer et~al\mbox{.}(2023)]%
        {schemmer_appropriate_2023}
\bibfield{author}{\bibinfo{person}{Max Schemmer}, \bibinfo{person}{Niklas Kuehl}, \bibinfo{person}{Carina Benz}, \bibinfo{person}{Andrea Bartos}, {and} \bibinfo{person}{Gerhard Satzger}.} \bibinfo{year}{2023}\natexlab{}.
\newblock \showarticletitle{Appropriate reliance on {AI} advice: conceptualization and the effect of explanations}. In \bibinfo{booktitle}{\emph{Proceedings of the 28th {International} {Conference} on {Intelligent} {User} {Interfaces}}} \emph{(\bibinfo{series}{{IUI} '23})}. \bibinfo{publisher}{ACM}, \bibinfo{address}{Sydney, NSW, Australia}, \bibinfo{pages}{410--422}.
\newblock
\showISBNx{9798400701061}
\urldef\tempurl%
\url{https://doi.org/10.1145/3581641.3584066}
\showDOI{\tempurl}


\bibitem[Schlögl et~al\mbox{.}(2019)]%
        {schlogl_artificial_2019}
\bibfield{author}{\bibinfo{person}{Stephan Schlögl}, \bibinfo{person}{Claudia Postulka}, \bibinfo{person}{Reinhard Bernsteiner}, {and} \bibinfo{person}{Christian Ploder}.} \bibinfo{year}{2019}\natexlab{}.
\newblock \showarticletitle{Artificial intelligence tool penetration in business: adoption, challenges and fears}. In \bibinfo{booktitle}{\emph{Knowledge {Management} in {Organizations}}} \emph{(\bibinfo{series}{{KMO} 2019})}. \bibinfo{publisher}{Springer International Publishing}, \bibinfo{address}{Zamora, Spain}, \bibinfo{pages}{259--270}.
\newblock
\showISBNx{978-3-030-21451-7}
\urldef\tempurl%
\url{https://doi.org/10.1007/978-3-030-21451-7_22}
\showDOI{\tempurl}


\bibitem[Schmidt and Biessmann(2020)]%
        {schmidt_calibrating_2020}
\bibfield{author}{\bibinfo{person}{Philipp Schmidt} {and} \bibinfo{person}{Felix Biessmann}.} \bibinfo{year}{2020}\natexlab{}.
\newblock \showarticletitle{Calibrating human-{AI} collaboration: impact of risk, ambiguity and transparency on algorithmic bias}. In \bibinfo{booktitle}{\emph{Machine {Learning} and {Knowledge} {Extraction}}} \emph{(\bibinfo{series}{{CD}-{MAKE} 2020})}. \bibinfo{publisher}{Springer International Publishing}, \bibinfo{address}{Dublin, Ireland}, \bibinfo{pages}{431--449}.
\newblock
\showISBNx{978-3-030-57320-1 978-3-030-57321-8}
\urldef\tempurl%
\url{https://doi.org/10.1007/978-3-030-57321-8_24}
\showDOI{\tempurl}


\bibitem[Shappell et~al\mbox{.}(2007)]%
        {shappell_human_2007}
\bibfield{author}{\bibinfo{person}{Scott Shappell}, \bibinfo{person}{Cristy Detwiler}, \bibinfo{person}{Kali Holcomb}, \bibinfo{person}{Carla Hackworth}, \bibinfo{person}{Albert Boquet}, {and} \bibinfo{person}{Douglas~A. Wiegmann}.} \bibinfo{year}{2007}\natexlab{}.
\newblock \showarticletitle{Human error and commercial aviation accidents: an analysis using the human factors analysis and classification system}.
\newblock \bibinfo{journal}{\emph{Human Factors: The Journal of the Human Factors and Ergonomics Society}} \bibinfo{volume}{49}, \bibinfo{number}{2} (\bibinfo{date}{April} \bibinfo{year}{2007}), \bibinfo{pages}{227--242}.
\newblock
\showISSN{0018-7208}
\urldef\tempurl%
\url{https://doi.org/10.1518/001872007X312469}
\showDOI{\tempurl}
\newblock
\shownote{Publisher: SAGE Publications Inc}.


\bibitem[Shneiderman(2020)]%
        {shneiderman_design_2020}
\bibfield{author}{\bibinfo{person}{Ben Shneiderman}.} \bibinfo{year}{2020}\natexlab{}.
\newblock \showarticletitle{Design lessons from {AI}'s two grand goals: human emulation and useful applications}.
\newblock \bibinfo{journal}{\emph{IEEE Transactions on Technology and Society}} \bibinfo{volume}{1}, \bibinfo{number}{2} (\bibinfo{date}{June} \bibinfo{year}{2020}), \bibinfo{pages}{73--82}.
\newblock
\showISSN{2637-6415}
\urldef\tempurl%
\url{https://doi.org/10.1109/TTS.2020.2992669}
\showDOI{\tempurl}


\bibitem[Sivaraman et~al\mbox{.}(2023)]%
        {sivaraman_ignore_2023}
\bibfield{author}{\bibinfo{person}{Venkatesh Sivaraman}, \bibinfo{person}{Leigh~A. Bukowski}, \bibinfo{person}{Joel Levin}, \bibinfo{person}{Jeremy~M. Kahn}, {and} \bibinfo{person}{Adam Perer}.} \bibinfo{year}{2023}\natexlab{}.
\newblock \showarticletitle{Ignore, trust, or negotiate: understanding clinician acceptance of {AI}-based treatment recommendations in health care}. In \bibinfo{booktitle}{\emph{Proceedings of the 2023 {CHI} {Conference} on {Human} {Factors} in {Computing} {Systems}}} \emph{(\bibinfo{series}{{CHI} '23})}. \bibinfo{publisher}{ACM}, \bibinfo{address}{Hamburg, Germany}, \bibinfo{pages}{754:1--754:18}.
\newblock
\urldef\tempurl%
\url{https://doi.org/10.1145/3544548.3581075}
\showDOI{\tempurl}


\bibitem[Smith et~al\mbox{.}(1997)]%
        {smith_brittleness_1997}
\bibfield{author}{\bibinfo{person}{P.J. Smith}, \bibinfo{person}{C.E. McCoy}, {and} \bibinfo{person}{C. Layton}.} \bibinfo{year}{1997}\natexlab{}.
\newblock \showarticletitle{Brittleness in the design of cooperative problem-solving systems: the effects on user performance}.
\newblock \bibinfo{journal}{\emph{IEEE Transactions on Systems, Man, and Cybernetics - Part A: Systems and Humans}} \bibinfo{volume}{27}, \bibinfo{number}{3} (\bibinfo{date}{May} \bibinfo{year}{1997}), \bibinfo{pages}{360--371}.
\newblock
\showISSN{1558-2426}
\urldef\tempurl%
\url{https://doi.org/10.1109/3468.568744}
\showDOI{\tempurl}


\bibitem[Smith-Renner et~al\mbox{.}(2020)]%
        {smith-renner_no_2020}
\bibfield{author}{\bibinfo{person}{Alison Smith-Renner}, \bibinfo{person}{Ron Fan}, \bibinfo{person}{Melissa Birchfield}, \bibinfo{person}{Tongshuang Wu}, \bibinfo{person}{Jordan Boyd-Graber}, \bibinfo{person}{Daniel~S. Weld}, {and} \bibinfo{person}{Leah Findlater}.} \bibinfo{year}{2020}\natexlab{}.
\newblock \showarticletitle{No explainability without accountability: an empirical study of explanations and feedback in interactive {ML}}. In \bibinfo{booktitle}{\emph{Proceedings of the 2020 {CHI} {Conference} on {Human} {Factors} in {Computing} {Systems}}} \emph{(\bibinfo{series}{{CHI} '20})}. \bibinfo{publisher}{Association for Computing Machinery}, \bibinfo{address}{Honolulu, HI, USA}, \bibinfo{pages}{497:1--497:13}.
\newblock
\showISBNx{978-1-4503-6708-0}
\urldef\tempurl%
\url{https://doi.org/10.1145/3313831.3376624}
\showDOI{\tempurl}


\bibitem[van Berkel et~al\mbox{.}(2021)]%
        {van_berkel_designing_2021}
\bibfield{author}{\bibinfo{person}{Niels van Berkel}, \bibinfo{person}{Omer~F. Ahmad}, \bibinfo{person}{Danail Stoyanov}, \bibinfo{person}{Laurence Lovat}, {and} \bibinfo{person}{Ann Blandford}.} \bibinfo{year}{2021}\natexlab{}.
\newblock \showarticletitle{Designing visual markers for continuous artificial intelligence support: a colonoscopy case study}.
\newblock \bibinfo{journal}{\emph{ACM Transactions on Computing for Healthcare}} \bibinfo{volume}{2}, \bibinfo{number}{1} (\bibinfo{date}{Dec.} \bibinfo{year}{2021}), \bibinfo{pages}{7:1--7:24}.
\newblock
\urldef\tempurl%
\url{https://doi.org/10.1145/3422156}
\showDOI{\tempurl}


\bibitem[Varghese(2020)]%
        {varghese_artificial_2020}
\bibfield{author}{\bibinfo{person}{Julian Varghese}.} \bibinfo{year}{2020}\natexlab{}.
\newblock \showarticletitle{Artificial intelligence in medicine: chances and challenges for wide clinical adoption}.
\newblock \bibinfo{journal}{\emph{Visceral Medicine}} \bibinfo{volume}{36}, \bibinfo{number}{6} (\bibinfo{date}{Oct.} \bibinfo{year}{2020}), \bibinfo{pages}{443--449}.
\newblock
\showISSN{2297-4725}
\urldef\tempurl%
\url{https://doi.org/10.1159/000511930}
\showDOI{\tempurl}


\bibitem[Vasconcelos et~al\mbox{.}(2023)]%
        {vasconcelos_explanations_2023}
\bibfield{author}{\bibinfo{person}{Helena Vasconcelos}, \bibinfo{person}{Matthew Jörke}, \bibinfo{person}{Madeleine Grunde-McLaughlin}, \bibinfo{person}{Tobias Gerstenberg}, \bibinfo{person}{Michael~S. Bernstein}, {and} \bibinfo{person}{Ranjay Krishna}.} \bibinfo{year}{2023}\natexlab{}.
\newblock \showarticletitle{Explanations can reduce overreliance on {AI} systems during decision-making}.
\newblock \bibinfo{journal}{\emph{Proceedings of the ACM on Human-Computer Interaction}} \bibinfo{volume}{7}, \bibinfo{number}{CSCW1} (\bibinfo{date}{April} \bibinfo{year}{2023}), \bibinfo{pages}{129:1--129:38}.
\newblock
\urldef\tempurl%
\url{https://doi.org/10.1145/3579605}
\showDOI{\tempurl}


\bibitem[Wang et~al\mbox{.}(2019)]%
        {wang_designing_2019}
\bibfield{author}{\bibinfo{person}{Danding Wang}, \bibinfo{person}{Qian Yang}, \bibinfo{person}{Ashraf Abdul}, {and} \bibinfo{person}{Brian~Y. Lim}.} \bibinfo{year}{2019}\natexlab{}.
\newblock \showarticletitle{Designing theory-driven user-centric explainable {AI}}. In \bibinfo{booktitle}{\emph{Proceedings of the 2019 {CHI} {Conference} on {Human} {Factors} in {Computing} {Systems}}} \emph{(\bibinfo{series}{{CHI} '19})}. \bibinfo{publisher}{ACM}, \bibinfo{address}{Glasgow, Scotland, UK}, \bibinfo{pages}{601:1--601:15}.
\newblock
\showISBNx{978-1-4503-5970-2}
\urldef\tempurl%
\url{https://doi.org/10.1145/3290605.3300831}
\showDOI{\tempurl}


\bibitem[Wang and Yin(2021)]%
        {wang_are_2021}
\bibfield{author}{\bibinfo{person}{Xinru Wang} {and} \bibinfo{person}{Ming Yin}.} \bibinfo{year}{2021}\natexlab{}.
\newblock \showarticletitle{Are explanations helpful? {A} comparative study of the effects of explanations in {AI}-assisted decision-making}. In \bibinfo{booktitle}{\emph{Proceedings of the 26th {International} {Conference} on {Intelligent} {User} {Interfaces}}} \emph{(\bibinfo{series}{{IUI} '21})}. \bibinfo{publisher}{ACM}, \bibinfo{address}{College Station, TX, USA}, \bibinfo{pages}{318--328}.
\newblock
\urldef\tempurl%
\url{https://doi.org/10.1145/3397481.3450650}
\showDOI{\tempurl}


\bibitem[Woods(1986)]%
        {woods_paradigms_1986}
\bibfield{author}{\bibinfo{person}{David~D. Woods}.} \bibinfo{year}{1986}\natexlab{}.
\newblock \showarticletitle{Paradigms for intelligent decision support}. In \bibinfo{booktitle}{\emph{Intelligent {Decision} {Support} in {Process} {Environments}}} \emph{(\bibinfo{series}{{NATO} {ASI} {Series}}, Vol.~\bibinfo{volume}{21})}. \bibinfo{publisher}{Springer Berlin, Heidelberg}, \bibinfo{address}{San Miniato, Italy}, \bibinfo{pages}{153--173}.
\newblock
\showISBNx{978-3-642-50331-3 978-3-642-50329-0}
\urldef\tempurl%
\url{https://doi.org/10.1007/978-3-642-50329-0_11}
\showDOI{\tempurl}


\bibitem[Würfel et~al\mbox{.}(2023)]%
        {wurfel_intelligent_2023}
\bibfield{author}{\bibinfo{person}{Jakob Würfel}, \bibinfo{person}{Boris Djartov}, \bibinfo{person}{Anne Papenfuß}, {and} \bibinfo{person}{Matthias Wies}.} \bibinfo{year}{2023}\natexlab{}.
\newblock \showarticletitle{Intelligent {Pilot} {Advisory} {System}: {The} journey from ideation to an early system design of an {AI}-based decision support system for airline flight decks}. In \bibinfo{booktitle}{\emph{Human {Factors} in {Transportation}}} \emph{(\bibinfo{series}{{AHFE} 2023}, Vol.~\bibinfo{volume}{95})}. \bibinfo{publisher}{AHFE Open Acces}, \bibinfo{address}{San Francisco, CA, USA}, \bibinfo{pages}{589--597}.
\newblock
\urldef\tempurl%
\url{https://doi.org/10.54941/ahfe1003844}
\showDOI{\tempurl}


\bibitem[Yang et~al\mbox{.}(2023)]%
        {yang_harnessing_2023}
\bibfield{author}{\bibinfo{person}{Qian Yang}, \bibinfo{person}{Yuexing Hao}, \bibinfo{person}{Kexin Quan}, \bibinfo{person}{Stephen Yang}, \bibinfo{person}{Yiran Zhao}, \bibinfo{person}{Volodymyr Kuleshov}, {and} \bibinfo{person}{Fei Wang}.} \bibinfo{year}{2023}\natexlab{}.
\newblock \showarticletitle{Harnessing biomedical literature to calibrate clinicians’ trust in {AI} decision support systems}. In \bibinfo{booktitle}{\emph{Proceedings of the 2023 {CHI} {Conference} on {Human} {Factors} in {Computing} {Systems}}} \emph{(\bibinfo{series}{{CHI} '23})}. \bibinfo{publisher}{ACM}, \bibinfo{address}{Hamburg, Germany}, \bibinfo{pages}{14:1--14:14}.
\newblock
\showISBNx{978-1-4503-9421-5}
\urldef\tempurl%
\url{https://doi.org/10.1145/3544548.3581393}
\showDOI{\tempurl}


\bibitem[Yang et~al\mbox{.}(2019)]%
        {yang_unremarkable_2019}
\bibfield{author}{\bibinfo{person}{Qian Yang}, \bibinfo{person}{Aaron Steinfeld}, {and} \bibinfo{person}{John Zimmerman}.} \bibinfo{year}{2019}\natexlab{}.
\newblock \showarticletitle{Unremarkable {AI}: fitting intelligent decision support into critical, clinical decision-making processes}. In \bibinfo{booktitle}{\emph{Proceedings of the 2019 {CHI} {Conference} on {Human} {Factors} in {Computing} {Systems}}} \emph{(\bibinfo{series}{{CHI} '19})}. \bibinfo{publisher}{ACM}, \bibinfo{address}{Glasgow, Scotland, UK}, \bibinfo{pages}{238:1--238:11}.
\newblock
\showISBNx{978-1-4503-5970-2}
\urldef\tempurl%
\url{https://doi.org/10.1145/3290605.3300468}
\showDOI{\tempurl}


\bibitem[Yang et~al\mbox{.}(2016)]%
        {yang_investigating_2016}
\bibfield{author}{\bibinfo{person}{Qian Yang}, \bibinfo{person}{John Zimmerman}, \bibinfo{person}{Aaron Steinfeld}, \bibinfo{person}{Lisa Carey}, {and} \bibinfo{person}{James~F. Antaki}.} \bibinfo{year}{2016}\natexlab{}.
\newblock \showarticletitle{Investigating the heart pump implant decision process: opportunities for decision support tools to help}. In \bibinfo{booktitle}{\emph{Proceedings of the 2016 {CHI} {Conference} on {Human} {Factors} in {Computing} {Systems}}} \emph{(\bibinfo{series}{{CHI} '16})}. \bibinfo{publisher}{ACM}, \bibinfo{address}{San Jose, CA, USA}, \bibinfo{pages}{4477--4488}.
\newblock
\urldef\tempurl%
\url{https://doi.org/10.1145/2858036.2858373}
\showDOI{\tempurl}


\bibitem[Yu et~al\mbox{.}(2018)]%
        {yu_artificial_2018}
\bibfield{author}{\bibinfo{person}{Kun-Hsing Yu}, \bibinfo{person}{Andrew~L. Beam}, {and} \bibinfo{person}{Isaac~S. Kohane}.} \bibinfo{year}{2018}\natexlab{}.
\newblock \showarticletitle{Artificial intelligence in healthcare}.
\newblock \bibinfo{journal}{\emph{Nature Biomedical Engineering}} \bibinfo{volume}{2}, \bibinfo{number}{10} (\bibinfo{date}{Oct.} \bibinfo{year}{2018}), \bibinfo{pages}{719--731}.
\newblock
\showISSN{2157-846X}
\urldef\tempurl%
\url{https://doi.org/10.1038/s41551-018-0305-z}
\showDOI{\tempurl}
\newblock
\shownote{Publisher: Nature Publishing Group}.


\bibitem[Zhang et~al\mbox{.}(2024)]%
        {zhang_rethinking_2024}
\bibfield{author}{\bibinfo{person}{Shao Zhang}, \bibinfo{person}{Jianing Yu}, \bibinfo{person}{Xuhai Xu}, \bibinfo{person}{Changchang Yin}, \bibinfo{person}{Yuxuan Lu}, \bibinfo{person}{Bingsheng Yao}, \bibinfo{person}{Melanie Tory}, \bibinfo{person}{Lace~M. Padilla}, \bibinfo{person}{Jeffrey Caterino}, \bibinfo{person}{Ping Zhang}, {and} \bibinfo{person}{Dakuo Wang}.} \bibinfo{year}{2024}\natexlab{}.
\newblock \showarticletitle{Rethinking human-{AI} collaboration in complex medical decision making: a case study in sepsis diagnosis}. In \bibinfo{booktitle}{\emph{Proceedings of the 2024 {CHI} {Conference} on {Human} {Factors} in {Computing} {Systems}}} \emph{(\bibinfo{series}{{CHI} '24})}. \bibinfo{publisher}{ACM}, \bibinfo{address}{Honolulu, HI, USA}, \bibinfo{pages}{445:1--445:18}.
\newblock
\showISBNx{9798400703300}
\urldef\tempurl%
\url{https://doi.org/10.1145/3613904.3642343}
\showDOI{\tempurl}


\bibitem[Zhang et~al\mbox{.}(2020)]%
        {zhang_effect_2020}
\bibfield{author}{\bibinfo{person}{Yunfeng Zhang}, \bibinfo{person}{Q.~Vera Liao}, {and} \bibinfo{person}{Rachel K.~E. Bellamy}.} \bibinfo{year}{2020}\natexlab{}.
\newblock \showarticletitle{Effect of confidence and explanation on accuracy and trust calibration in {AI}-assisted decision making}. In \bibinfo{booktitle}{\emph{Proceedings of the 2020 {Conference} on {Fairness}, {Accountability}, and {Transparency}}} \emph{(\bibinfo{series}{{FAT}* '20})}. \bibinfo{publisher}{ACM}, \bibinfo{address}{Barcelona, Spain}, \bibinfo{pages}{295--305}.
\newblock
\showISBNx{978-1-4503-6936-7}
\urldef\tempurl%
\url{https://doi.org/10.1145/3351095.3372852}
\showDOI{\tempurl}


\bibitem[Zhang et~al\mbox{.}(2021)]%
        {zhang_forward_2021}
\bibfield{author}{\bibinfo{person}{Zelun~Tony Zhang}, \bibinfo{person}{Yuanting Liu}, {and} \bibinfo{person}{Heinrich Hussmann}.} \bibinfo{year}{2021}\natexlab{}.
\newblock \showarticletitle{Forward reasoning decision support: toward a more complete view of the human-{AI} interaction design space}. In \bibinfo{booktitle}{\emph{{CHItaly} 2021: 14th {Biannual} {Conference} of the {Italian} {SIGCHI} {Chapter}}} \emph{(\bibinfo{series}{{CHItaly} '21})}. \bibinfo{publisher}{ACM}, \bibinfo{address}{Bolzano, Italy}, \bibinfo{pages}{18:1--18:5}.
\newblock
\showISBNx{978-1-4503-8977-8}
\urldef\tempurl%
\url{https://doi.org/10.1145/3464385.3464696}
\showDOI{\tempurl}


\bibitem[Zhang et~al\mbox{.}(2023)]%
        {zhang_resilience_2023}
\bibfield{author}{\bibinfo{person}{Zelun~Tony Zhang}, \bibinfo{person}{Cara Storath}, \bibinfo{person}{Yuanting Liu}, {and} \bibinfo{person}{Andreas Butz}.} \bibinfo{year}{2023}\natexlab{}.
\newblock \showarticletitle{Resilience through appropriation: pilots' view on complex decision support}. In \bibinfo{booktitle}{\emph{Proceedings of the 28th {International} {Conference} on {Intelligent} {User} {Interfaces}}} \emph{(\bibinfo{series}{{IUI} '23})}. \bibinfo{publisher}{ACM}, \bibinfo{address}{Sydney, NSW, Australia}, \bibinfo{pages}{397--409}.
\newblock
\showISBNx{979-8-4007-0106-1/23/03}
\urldef\tempurl%
\url{https://doi.org/10.1145/3581641.3584056}
\showDOI{\tempurl}


\end{thebibliography}

\appendix

\section{Exit Interview Guide}
\label{sec:interview_guide}
\begin{itemize}
    \item How was your impression of the system?
    \item On a scale from 1---\textit{not helpful at all} to 5---\textit{extremely helpful}, how do you rate the helpfulness of the system?
    \item What is the greatest added value of the system?
    \item What do you find problematic or in need of improvement about the system?
    \item What was your strategy for using the system? What did you use it for?
    \item \textit{For Rec, Cont, Rec+Cont conditions}: How important was/were the recommendations/normal flight mode/color highlights for your usage?
    \item \textit{For Baseline condition}: What kind of AI features would you like to see in the system?
    \item Anything else that you want to ask or comment on?
\end{itemize}

\section{Participant Details}
\label{sec:participant_details}
\begin{table}[thb]
    \caption{Focus group participant details.}
    \label{tab:focus_participants}
    \centering
    \begin{tabular}{r r r r c}
        \toprule
        ID & Gender & Rank & Flight hours & Participated in pilot test \\
        \midrule
        F1 & Male & F/O & 400 &  \\
        F2 & Male & Captain & \num{10200} &  \\
        F3 & Male & F/O & 1600 & X \\
        F4 & Male & F/O & 370 & X \\
        \midrule
        \makecell[tr]{\textit{Median}\\\textit{(IQR)}} & & & \makecell[tr]{1000\\(392.5--3750)} & \\
        \bottomrule
        \end{tabular}
\end{table}

\begin{table}[thb]
    \caption{Study participant details.}
    \label{tab:study_participants}
    \centering
    \begin{tabular}{r r r r c} 
        \toprule
        ID & Gender & Rank & Flight hours & \# Diversions \\
        \midrule
        B1 & Male & F/O & 3180 & 0 \\
        B2 & Male & F/O & 4000 & 4 \\
        B3 & Male & F/O & 2000 & 4 \\
        B4 & Male & F/O & 2000 & 1 \\
        B5 & Male & F/O & 200 & 0 \\
        B6 & Male & F/O & 1700 & 2 \\
        B7 & Male & F/O & 4800 & 3 \\
        B8 & Male & F/O & 5000 & 1 \\
        R1 & Male & F/O & 4000 & 2 \\
        R2 & Male & F/O & 2000 & 1 \\
        R3 & Male & F/O & 2500 & 1 \\
        R4 & Male & F/O & 4000 & 5 \\
        R5 & Male & F/O & 3000 & 0 \\
        R6 & Male & F/O & 1640 & 0 \\
        R7 & Male & Captain & \num{11000} & 1 \\
        R8 & Male & F/O & 2500 & 1 \\
        C1 & Male & F/O & 2400 & 0 \\
        C2 & Male & F/O & 3000 & 0 \\
        C3 & Male & F/O & 6000 & 7 \\
        C4 & Male & F/O & 2500 & 2 \\
        C5 & Male & Captain & \num{13000} & 5 \\
        C6 & Male & F/O & 2500 & 0 \\
        C7 & Male & F/O & 4000 & 5 \\
        C8 & Female & F/O & 580 & 0 \\
        RC1 & Male & F/O & 3800 & 3 \\
        RC2 & Male & F/O & 4000 & 0 \\
        RC3 & Male & F/O & 2000 & 3 \\
        RC4 & Male & F/O & 2400 & 2 \\
        RC5 & Male & F/O & 1900 & 3 \\
        RC6 & Male & F/O & 800 & 0 \\
        RC7 & Female & F/O & 2500 & 2 \\
        RC8 & Male & F/O & 3000 & 2 \\
        \midrule
        \makecell[tr]{\textit{Median}\\\textit{(IQR)}} & & & \makecell[tr]{2500\\(2000--4000)} & \makecell[tc]{1.5\\(0--3)} \\
        \bottomrule
    \end{tabular}
\end{table}

\end{document}